\documentclass[preprint]{elsarticle}

\makeatletter
\def\ps@pprintTitle{%
 \let\@oddhead\@empty
 \let\@evenhead\@empty
 \def\@oddfoot{\centerline{\thepage}}%
 \let\@evenfoot\@oddfoot}
\makeatother

\usepackage{lineno,hyperref}

\usepackage{booktabs} % For formal tables
\usepackage[ruled,vlined,linesnumbered]{algorithm2e}
\usepackage{algorithmic}
\usepackage{adjustbox}
\usepackage{enumitem}
\usepackage{xspace}
\usepackage{mdwlist}
\usepackage{url}
\usepackage{autobreak}
\usepackage{color}
\usepackage{tablefootnote} % For table footnote
\usepackage[center]{subfigure} % for \subfloat
\usepackage[mathscr]{euscript} % for euler script.
\usepackage{makecell}
\usepackage{footnote} % For footnotes in tables (\savenotes)
\usepackage{multirow}
\usepackage{tikz}
\usepackage{amsfonts}
\usepackage{lineno}

\newtheorem{problem}{Problem}

\newcommand{\hide}[1]{}

\newcommand{\subfloat}{\subfigure}

\newcommand{\method}{\textsc{GSHL}\xspace}

\DeclareMathOperator*{\argmin}{arg\,min}

\modulolinenumbers[5]

\begin{document}
\let\makeLineNumber\relax

\begin{frontmatter}

\title{Finding Key Structures in MMORPG Graph with Hierarchical Graph Summarization}

\author[1]{Jun-Gi Jang}
\author[1]{Chaeheum Park}
\author[2]{Changwon Jang}
\author[2]{Geonsoo Kim}
\author[1]{U Kang\corref{mycorrespondingauthor}}

\address[1]{Computer Science and Engineering, Seoul National University, Seoul, Republic of Korea}
\address[2]{Knowledge AI Lab., NCSOFT Co., Seongnam, Republic of Korea}

%\fntext[myfootnote]{Corresponding author (ukang@snu.ac.kr)}
\cortext[mycorrespondingauthor]{Corresponding author}
\ead{ukang@snu.ac.kr}


\begin{abstract}
What are the key structures existing in a large real-world MMORPG (Massively Multiplayer Online Role-Playing Game) graph?
How can we compactly summarize an MMORPG graph with hierarchical node labels, considering consistent substructures at different levels of hierarchy?
Recent MMORPGs generate complex interactions between entities inducing a heterogeneous graph where each entity has hierarchical labels.
Succinctly summarizing a heterogeneous MMORPG graph is crucial to better understand its structure; however it is a challenging task
since it needs to handle complex interactions and hierarchical labels efficiently.
Although there exist few methods to summarize a large-scale graph,
they do not deal with heterogeneous graphs with hierarchical node labels.

We propose \method, a novel method that summarizes a heterogeneous graph with hierarchical labels.
We formulate the encoding cost of hierarchical labels using MDL (Minimum Description Length).
\method exploits the formulation to identify and segment subgraphs, and discovers compact and consistent structures in the graph.
Experiments on a large real-world MMORPG graph with multi-million edges show that
\method is a useful and scalable tool for summarizing the graph,
finding important and interesting structures in the graph,
and finding similar users.
\end{abstract}

\begin{keyword}
Graph summarization \sep minimum description length \sep hierarchical label \sep
massively multiplayer online role-playing game
\end{keyword}

\end{frontmatter}

\linenumbers

\section{Introduction}
\label{sec:intro}

% Problem Definition
What does a large real-world MMORPG (Massively Multiplayer Online Role-Playing Game) graph look like?
What are the key structures existing in a large real-world MMORPG (Massively Multiplayer Online Role-Playing Game) graph?
How can we compactly summarize an MMORPG graph with hierarchical node labels, considering consistent substructures at different levels of hierarchy?
A large number of interactions between heterogeneous entities appear in MMORPGs, and they are often represented as a heterogeneous graph.
For example, a user has several characters, and each character with various equipment visits dungeons.
Moreover, each entity has hierarchical labels (i.e., character-dealer-destroyer, and equipment-weapon).
Summarizing a heterogeneous graph with hierarchical node labels is a crucial problem to understand its characteristics.
A better understanding helps us identify important structures and obtain meaningful insights. % in the graph.

% Limitation, Problem Difficulties
Several approaches have been proposed to summarize graphs using significant structures (vocabularies) commonly found in real-world graphs. % such as star, clique, bipartite core, and chain.
VoG~\cite{KoutraKVF14,SAM:SAM11267} is the first approach to summarize a homogeneous graph with vocabulary terms by compressing the graph in terms of Minimum Description Length (MDL) principle.
TimeCrunch~\cite{ShahKZGF15} extended VoG method to summarize dynamic graphs.
Both methods, however, fail to summarize a heterogeneous graph with hierarchical labels since they do not excogitate labels of nodes nor the hierarchy of labels.
%The major challenges are 1) to define meaningful structures with node labels, and 2) to deal with node labels and label hierarchy based on MDL principle.

% Main Ideas and Main Contributions

\begin{figure*}[t!]
	\centering
	\includegraphics[width=0.9\textwidth]{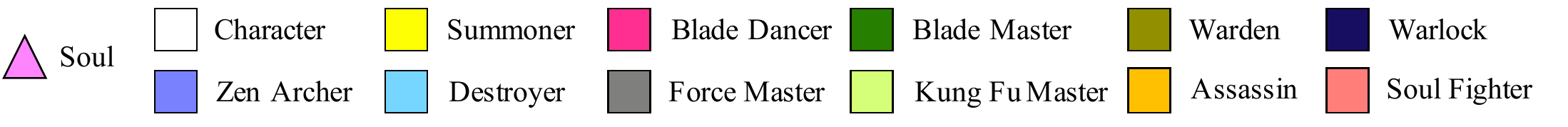} \\
	\includegraphics[width=0.9\textwidth]{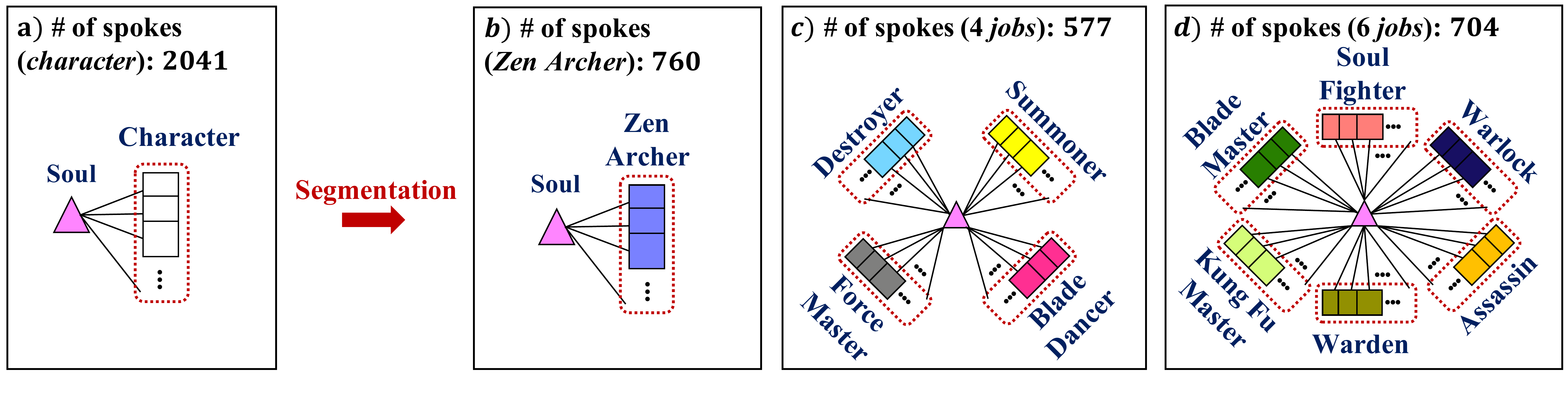}
%	 	 \subfloat[{\color{red}Fake figure}]{\includegraphics[width=0.25\textwidth]{FIG/CASE_SEGMENTATION.pdf}\label{fig:case}}\\
	\caption{
Our hierarchical graph summarization finds more precise and detailed graph structures, compared to non-hierarchical graph summarization.
(a) Non-hierarchical graph summarization only finds that an equipment is shared by many characters. It does not recognize the underlying substructures in the characters.
(b, c, d) Hierarchical graph summarization finds structures in the lower-level labels of the characters and
decomposes the character group into three major substructures with different subgroups.
%\blue{\textit{Zen archers} prefer the \textit{soul} \textit{equipment} (pink triangle) the most, compared to other characters. %, although any \textit{jobs} can have this \textit{equipment}.	}
}
	\label{fig:segmented}
\end{figure*}

\begin{figure}[t!]
	\centering
	\includegraphics[width=0.7\textwidth]{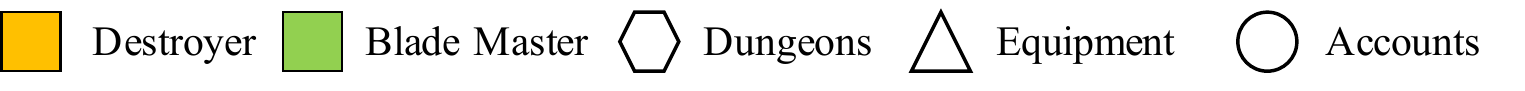} \\
	 	 \subfloat[The most similar account id $3602$]{
	 	 	\includegraphics[width=0.3\textwidth]{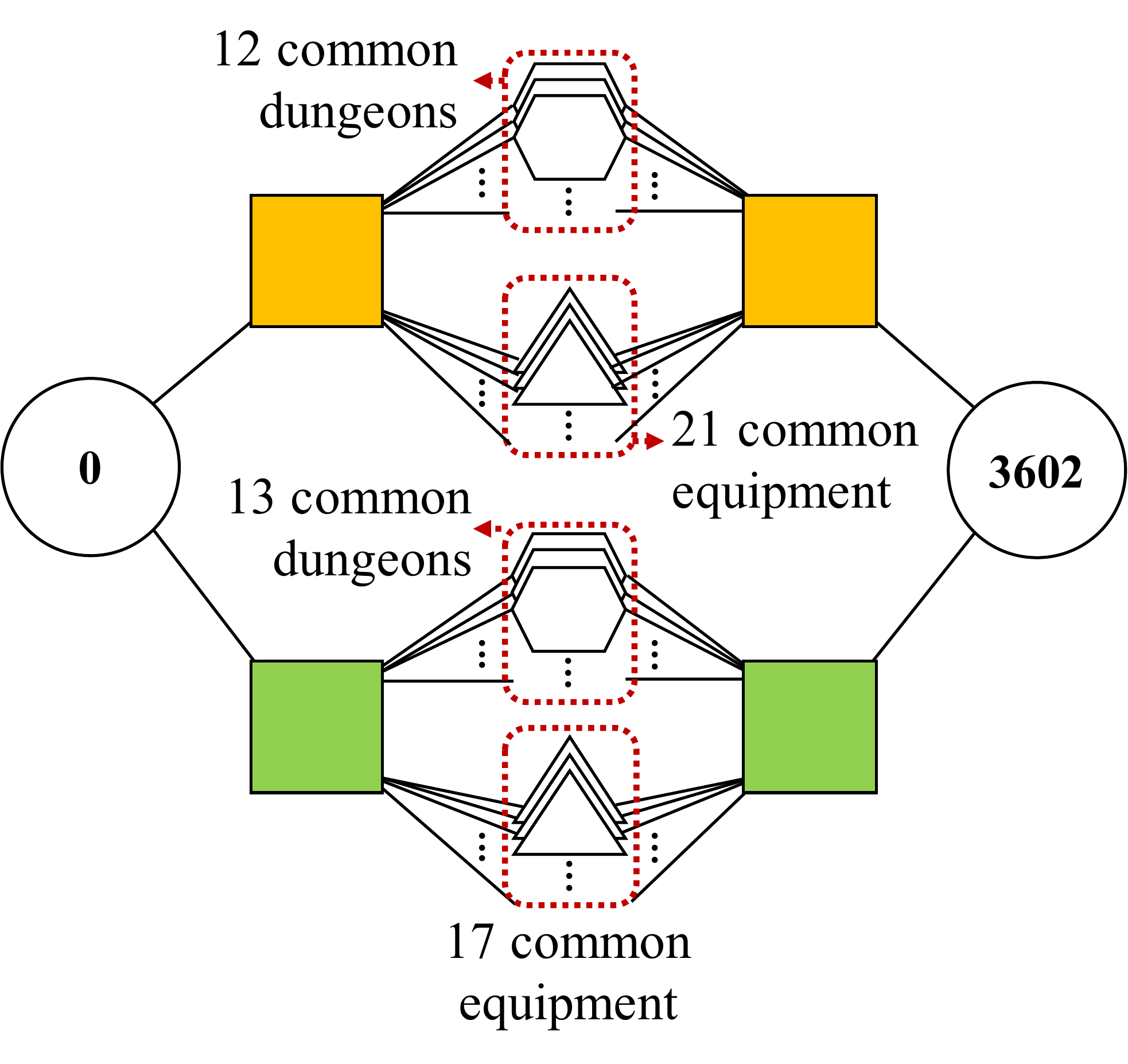}
	 	 	\label{fig:similar_account3602}}
	 	 \subfloat[The second most similar account id $9$]{
	 	 	\includegraphics[width=0.3\textwidth]{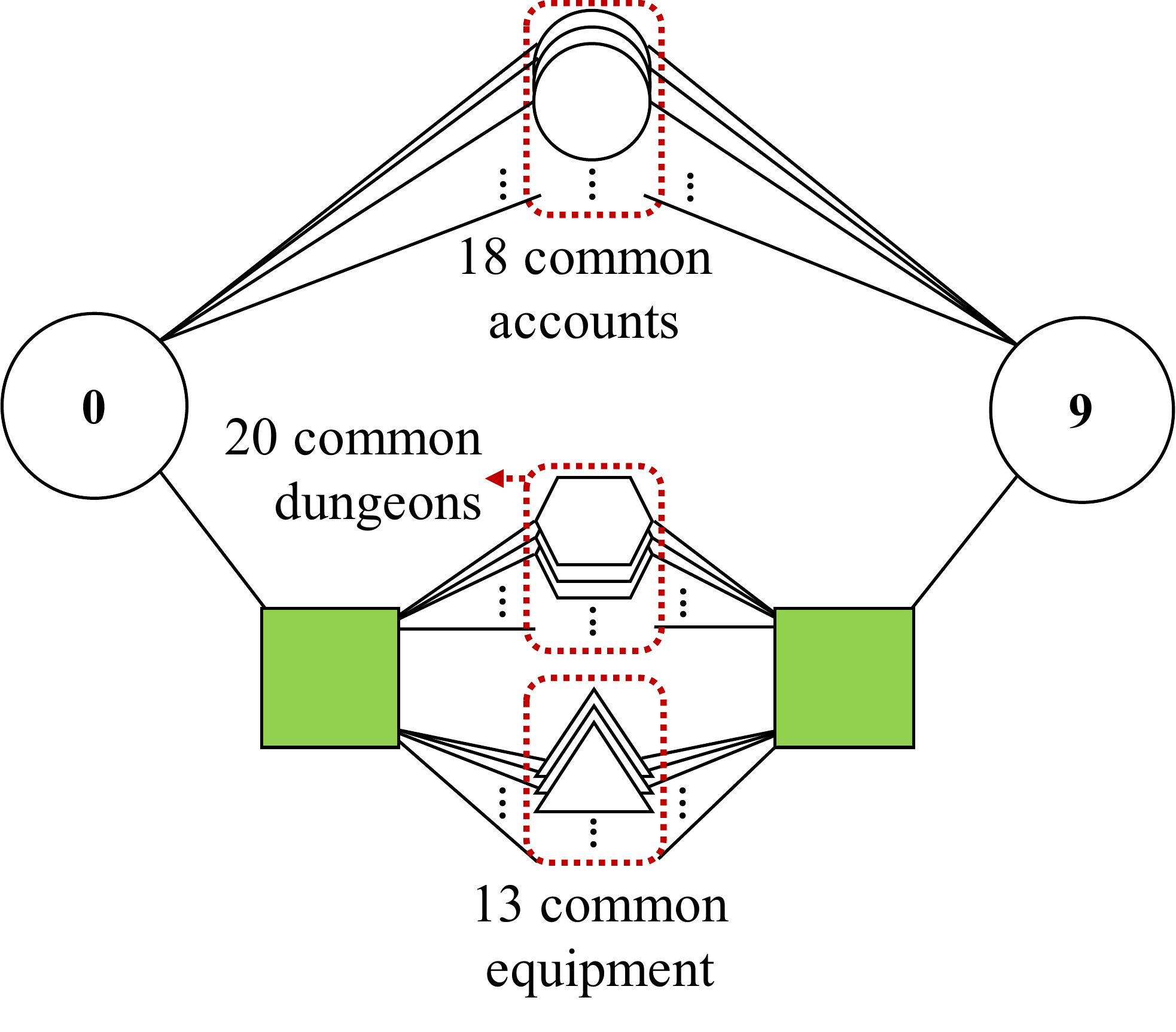}
	 	 	\label{fig:similar_account9}}
	\caption{The two most similar accounts for account id $0$ share common characters which have similar equipment and visited dungeons. They often share common friends as well.
}
%	 The number in the account node represents the id and the number in the boxes indicates the number of nodes it shares with the same label.}}
	\label{fig:similar}
\end{figure}

We propose \method (Graph Summarization with Hierarchical Labels), a summarization method for a heterogeneous graph with hierarchical node labels.
Based on the MDL principle that good compression leads to good summarization,
we precisely define the encoding cost for heterogeneous graphs with hierarchical node labels.
%and finds meaningful structures which minimizes the cost.
%We then design \method to summarize a heterogeneous graph with hierarchical node labels.
\method decomposes a heterogeneous graph into subgraphs using a graph decomposition method,
and
encodes each subgraph as a structure (e.g., star, clique, bipartite core, chain, etc.).
Then \method further segments each structure by considering hierarchical labels of nodes in the structure, and summarizes the graph using the encoding cost.
Thanks to \method,
we analyze a large real-world MMORPG graph and find its key structures as well as similar users.
%
% List of Contributions
Our contributions are as follows:
\begin{enumerate*}
	\item \textbf{Problem formulation.}
We formulate the problem of summarizing a heterogeneous graph with hierarchical labels.
We use a large real-world graph extracted from Blade \& Soul, a popular MMORPG played worldwide.
This is the first work in literature that summarizes a large real-world MMORPG graph, to the best of our knowledge.
	\item \textbf{Scalable Method.}
We propose \method, a novel algorithm to summarize a heterogeneous graph with hierarchical node labels.
\method carefully exploits MDL to encode hierarchical labels,
and generates a succinct summary by segmenting subgraphs at different levels of hierarchy.
\method is near-linear on the number of edges (see Fig.~\ref{fig:scalability}).
	\item \textbf{Experiment.}
Experiments on the real-world MMORPG graph reveals that \method is a useful tool for
succinctly summarizing the graph (see Table~\ref{tab:resultcost}).
\method discovers interesting patterns (see Fig.~\ref{fig:segmented} and \ref{fig:comprehension}), and similar users with similar structures (see Fig.~\ref{fig:similar}).
\end{enumerate*}

%The rest of this paper is organized as follows.
%We describe the details of Blade \& Soul data in Section~\ref{sec:data}.
%We describe our proposed MDL formulation for hierarchical graph summarization in Section~\ref{sec:problem}.
%Then we describe GSHL, our proposed graph summarization algorithm for heterogeneous graphs with hierarchical node labels in Section~\ref{sec:graphsummary}.
%We present experimental results in Section~\ref{sec:experiment}.
%After discussing related works in Section~\ref{sec:related},
%we conclude in Section~\ref{sec:conclusion}.
%%
%Table~\ref{tab:notation} shows the symbols used. % in this paper.

In the rest of this paper, we describe data,
MDL formulation,
summary generation,
experimental results,
related works,
and
conclusion.

%In Section~\ref{sec:data}, we present the details of Blade \& Soul data.
%Then, we describe MDL formulation in Section~\ref{sec:problem}, and how to summarize the heterogeneous graph in Section~\ref{sec:graphsummary}.
%After presenting experimental results in Section~\ref{sec:experiment} and related works in Section~\ref{sec:related}, we conclude in Section~\ref{sec:conclusion}.

\section{Data Description}
\label{sec:data}

%We describe the data that we studied in this paper.
Our target data is from Blade \& Soul (\url{https://www.bladeandsoul.com/en/}), a popular MMORPG
played in more than $60$ countries worldwide and earning revenue worth USD $75$ million in $2019$.
%We present a brief introduction of Blade \& Soul, the contents of the data, and how we constructed a graph from the data.

\subsection{Blade \& Soul}
\label{sec:bns}

%Like most MMORPGs,
Blade \& Soul entails multiple characters simultaneously interacting with each other in an open-world environment.
Each character carries a variety of equipment acquired throughout its playtime;
it enters various dungeons with other characters in order to acquire new equipment or items. % that enhances their existing equipment.
Such game design encourages users to repetitively visit dungeons. %, often referred to as \textit{farming}.

%In Blade \& Soul there are multiple jobs, and thus
A user associated with an account can possess multiple in-game characters with different jobs.
{{These jobs are divided into three types of \textit{play-styles}}}: \textit{dealer}, \textit{tanker}, and \textit{buffer}.
A \textit{dealer} mainly afflicts damage to the enemies of its party (a group of multiple characters), while a \textit{tanker} taunts enemies so the party members can freely afflict damage, and a \textit{buffer} aids other party members.
A party needs to have balanced types of jobs to successfully enter a dungeon and fight with enemies.

\begin{figure}[!t]
	\centering
	\includegraphics[width=0.6\textwidth]{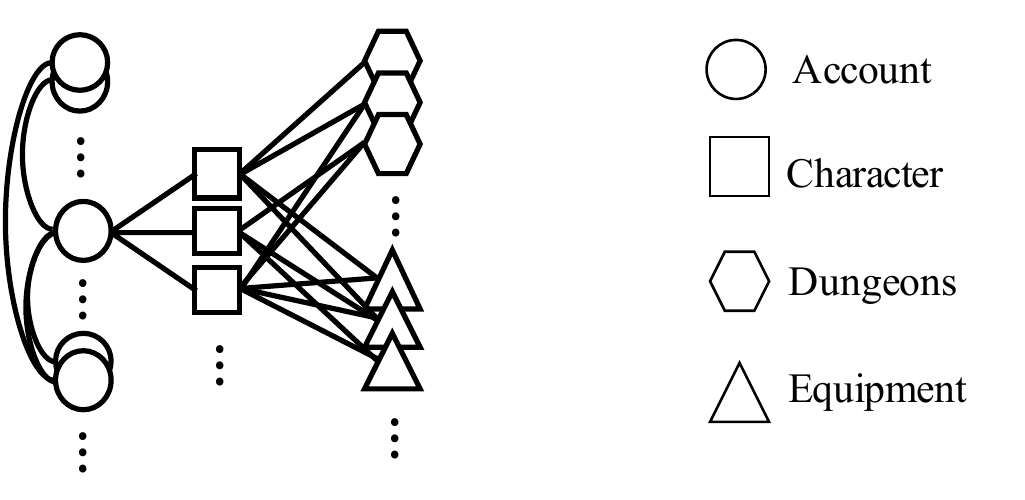} \\
	\caption{Graph structure of Blade \& Soul.
Accounts have characters which have equipment and dungeons they visited. Accounts are connected to each other via friendship.
}
	\label{fig:graph_form}
\end{figure}

\begin{figure*}[!t]
	\centering
	 	 	 \subfloat[Degree distribution of nodes for each level-$1$ label type]{\includegraphics[width=0.3\textwidth]{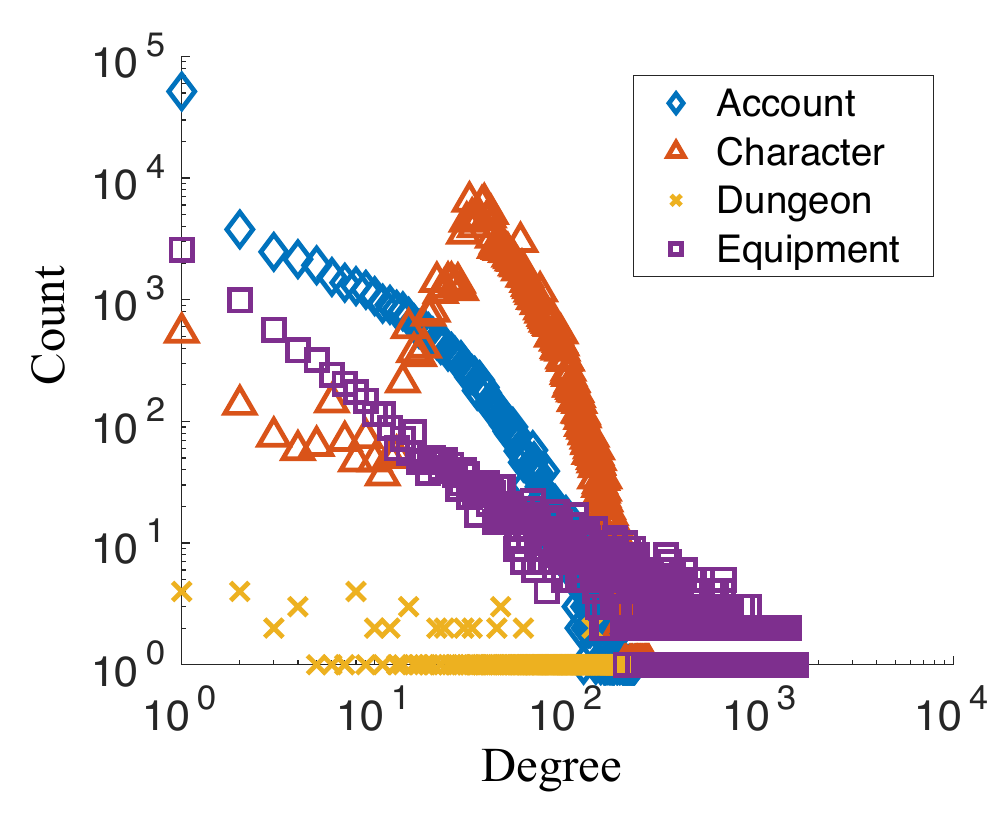}\label{fig:data_degree_type}}
	 \subfloat[Degree distribution of all nodes]{\includegraphics[width=0.3\textwidth]{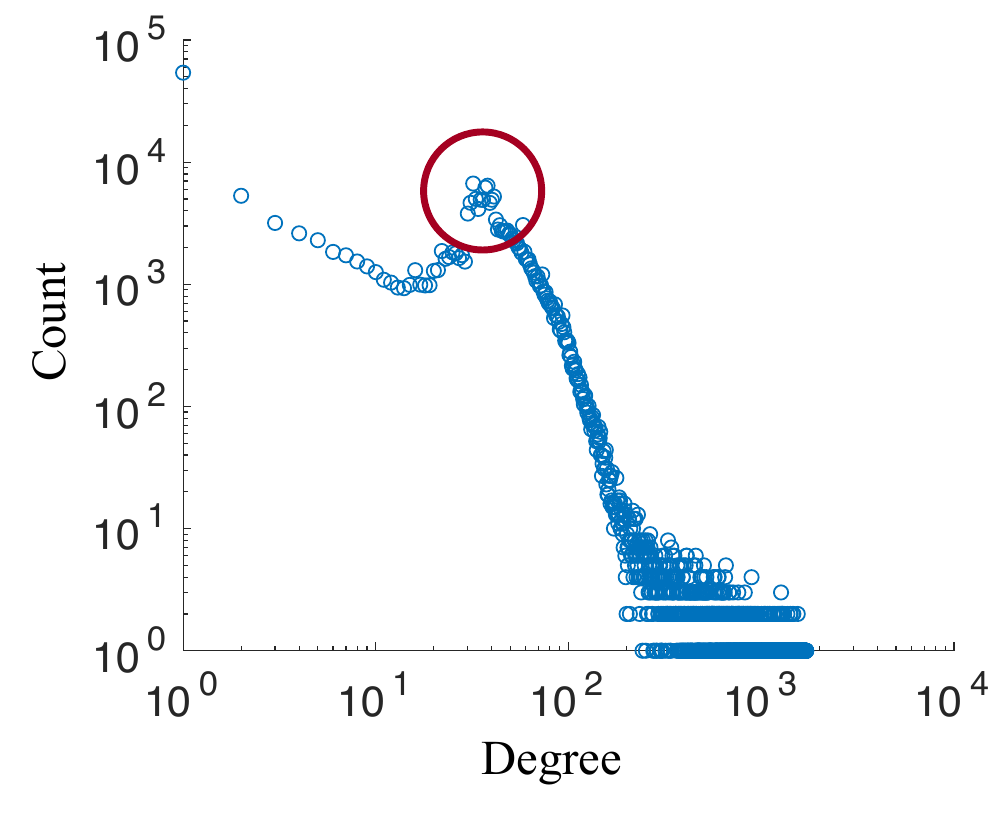}\label{fig:data_degree_all}}
	 	 \subfloat[Result of SlashBurn]{\includegraphics[width=0.25\textwidth]{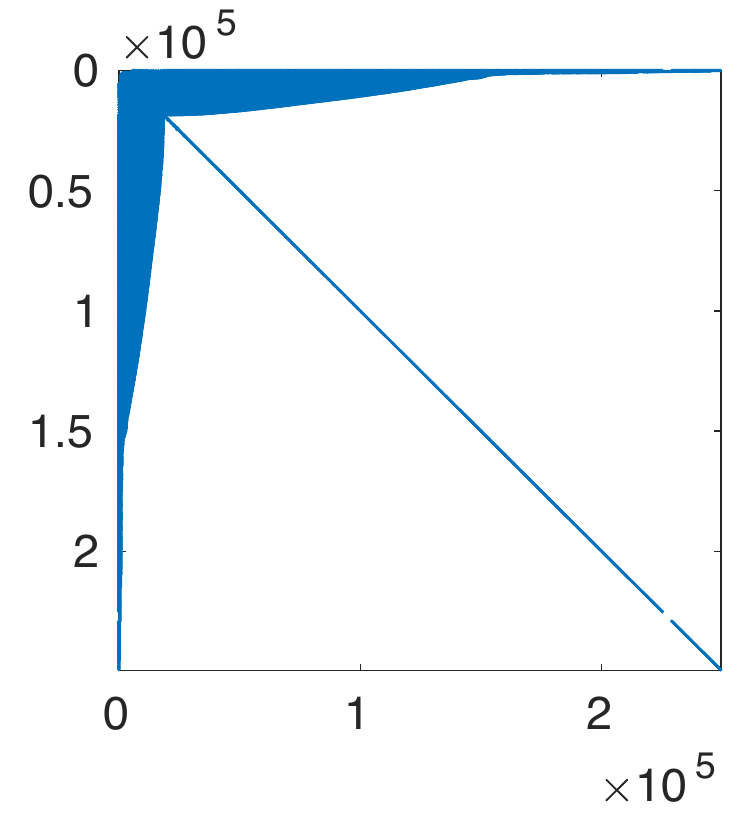}\label{fig:data_sb}}\\
	\caption{Blade \& Soul graph shows a skewed degree distribution.
}
	\label{fig:data_analysis}
\end{figure*}

\subsection{Dataset}

We collect $3$ months of data from January to March in 2019, and extract the following four relationships.

\begin{itemize}[noitemsep,topsep=0pt]
  \item \textit{account} - \textit{character}: which character(s) each account possesses.
  \item \textit{account} - \textit{account}: friendship between accounts.
  \item \textit{character} - \textit{dungeon}: which dungeons each character visited.
  \item \textit{character} - \textit{equipment}: which equipment each character is equipped with.
\end{itemize}

%From these relationships we construct a graph containing $249,455$ nodes and $7,885,487$ edges.
%
We extract $4$ types of entities: account, character, dungeon and equipment.
An account represents a user, and
a user can have multiple characters. % where each character belongs to one of the following jobs: force master, destroyer, summoner, blade dancer, zen archer, blade master, kung fu master, warden, assassin, warlock, and soul fighter.
%
%A dungeon entity has a difficulty of playing in it: normal, advanced, and others.
A dungeon entity belongs to a type based on its difficulty %: normal, advanced, and others,
where an advanced dungeon is a more difficult version of the corresponding normal dungeon.

\subsubsection{Hierarchical Labels}
Each label is further divided into sub-labels,
and this leads to a hierarchical label structure summarized in Table ~\ref{tab:label}.
The four types of entities (account, character, dungeon, and equipment)
extracted from the dataset represent level-$1$ of the hierarchy.
The account type has no sub-labels.
The character type has level-$2$ sub-labels based on \textit{play-styles}. %: \textit{dealer}, \textit{tanker}, and \textit{buffer}.
Each of the level-$2$ label of the character has level-$3$ sub-labels.
%\textit{Dealer} consists of force master, destroyer, summoner, blade dancer, and zen archer; \textit{tanker} consists of blade master, kung fu master, and warden; finally, \textit{buffer} consists of assassin, warlock, and soul fighter.
The dungeon type has $3$ level-$2$ sub-labels. %: normal, advanced and others.
The equipment type has $14$ level-$2$ sub-labels. %: weapon, soul shield, ring, bracelet, earring, belt, necklace, soul, heart, pet, glove, soul badge, mystic badge, and talisman.

\begin{table}[t]
\centering
\small
\caption{Hierarchical labels in Blade \& Soul graph.}
\label{tab:label}
%\resizebox{0.8\columnwidth}{!}{%
\begin{tabular}{cccr}
\toprule
\textbf{Level-1}                  & \textbf{Level-2}                 & \textbf{Level-3}   & \textbf{Node Count}     \\
\midrule
 \textbf{Account}                     & $\times$                & $\times$       &$83,970$\\ \hline
\multirow{11}{*}{\textbf{Character}} & \multirow{5}{*}{Dealer} & Force Master   	&$19,147$\\
                             &                         & Destroyer      &23,327\\
                             &                         & Summoner       &6,266\\
                             &                         & Blade Dancer   &5,822\\
                             &                         & Zen Archer     &11,023\\  \cline{2-3}
                             & \multirow{3}{*}{Tanker} & Blade Master   &11,854\\
                             &                         & Kung Fu Master &17,845\\
                             &                         & Warden         &6,689\\  \cline{2-3}
                             & \multirow{3}{*}{Buffer} & Assassin       &18,460\\
                             &                         & Warlock        &15,868\\
                             &                         & Soul Fighter   &18,249\\ \hline
 \multirow{3}{*}{\textbf{Dungeon}}    & Normal                  & $\times$       &154\\
                             & Advanced                 & $\times$       &12\\
                             & Others                   & $\times$      	&133 \\ \hline
 \multirow{14}{*}{\textbf{Equipment}} & Weapon                  & $\times$       &3,219\\
                             & Soul Shield             & $\times$    	&3,400\\
                             & Ring                    & $\times$       &431\\
                             & Bracelet                & $\times$       &398\\
                             & Earring                 & $\times$       &455\\
                             & Belt                    & $\times$       &95\\
                             & Necklace                & $\times$       &430\\
                             & Soul                    & $\times$       &171\\
                             & Heart                   & $\times$       &47\\
                             & Pet                     & $\times$       &269\\
                             & Glove                   & $\times$       &26\\
                             & Soul Badge              & $\times$     	&927\\
                             & Mystic Badge            & $\times$   	&739\\
                             & Talisman                & $\times$       &29\\
\midrule
\textbf{Total}&&&$249,455$\\
\bottomrule
\end{tabular}
\end{table}

\begin{table}[!t]
\centering
\small
\caption{Number of edges between level-1 labels in Blade \& Soul graph.}
\label{tab:edge}
%\resizebox{0.8\columnwidth}{!}{%
\begin{tabular}{lrrrr}
\toprule
          & \textbf{Account} & \textbf{Character} & \textbf{Dungeon} & \textbf{Equipment} \\
\midrule
\textbf{Account}   & 229,338  & 154,550    & 0       & 0         \\
\textbf{Character} & 154,550  & 0         & 2,680,520 & 4,821,079   \\
\textbf{Dungeon}   & 0       & 2,680,520   & 0       & 0         \\
\textbf{Equipment} & 0       & 4,821,079   & 0       & 0\\
\midrule
\textbf{Total}&&&&\textbf{7,885,487} \\
\bottomrule
\end{tabular}
\end{table}

\subsubsection{Graph}
An entity, which is either account, character, dungeon, or equipment, becomes a node in the graph.
A relationship between two entities forms an edge.
There are four types of relationships: 1) (account - account), 2) (account - character), 3) (character - dungeon), and 4) (character - equipment) as shown in Fig.~\ref{fig:graph_form}.
Table~\ref{tab:edge} represents the number of edges between level-$1$ labels. % in the graph.
The formed graph contains $249,455$ nodes and $7,885,487$ edges.% (see Tables \ref{tab:label} and \ref{tab:edge}).

Fig.~\ref{fig:data_analysis} shows the characteristics of the graph.
The degree distributions of nodes for each level-1 label type are shown in Fig.~\ref{fig:data_degree_type}.
Note that degree distributions of all node types except \textit{character} follow power-law.
The degree distribution of \textit{character} type nodes has a mode %(denoted by a red circle)
around the degrees $30 \sim 40$
due to the following reasons:
%1) a character is always associated with one account,
1) the number of items a character can have is limited, and 2) the number of dungeons a character can visit is limited.
Fig.~\ref{fig:data_degree_all} shows the degree distribution of all nodes.
Even though there is a mode around the degrees $30 \sim 40$, the graph shows a skewed degree distribution which is also verified in Fig.~\ref{fig:data_sb} which shows the reordered adjacency matrix from SlashBurn~\cite{KangF11,journals/tkde/LimKF14} method.
Note that a reordered adjacency matrix with a thin arrow shape like Fig.~\ref{fig:data_sb} means there are few high degree nodes in the graph which can be used for decomposing (or `shattering') it~\cite{KangF11}.

\section{Proposed Method: MDL Formulation}

%In this section, we explain our first contribution: formulation of graph summarization for heterogeneous graph with hierarchical labels, using MDL.

\subsection{Overview}

Our goal is to construct a concise summary of an MMORPG graph with hierarchical node labels.
There are several challenges for the goal.

\begin{enumerate*}
\item How can we summarize an MMORPG graph with a few key subgraphs (vocabularies)?
\item In designing an optimization objective for the summary, how can we incorporate hierarchical node labels?
\item Given a subgraph, how can we extract consistent substructures at different levels of hierarchy?
\end{enumerate*}

Our ideas to solve the challenges are as follows.

\begin{enumerate*}
\item \textbf{Exploiting Minimum Description Length (MDL)} allows us to extract key structures (e.g., clique, star, bipartite core, etc.) from the MMORPG graph.
%  (Sections~\ref{sec:mdl_problem} -~\ref{sec:mdl:connectivity}).
	We add the formulation for hierarchical labels while exploiting the formulation for structures used in VoG~\cite{KoutraKVF14}.
%Exploiting  (use MDL...) (It is similar to VoG, but we added ....) (Section III.B - III.D)
\item \textbf{Considering label consistency of structures} enables us to evaluate various structures in depth.
% (Sections~\ref{sec:hilabel} and~\ref{sec:error}).
We carefully formulate label consistency for each key structure.
%(Formulate how to encode hierarchical labels...) (Section III.E - III.F)
\item \textbf{Segmenting labels of an inconsistent structure from higher to lower levels} generates consistent substructures.
% (Section~\ref{subsec:structure_seg}).
By determining segmentation at each level of the hierarchy, we extract pivotal structures from a heterogeneous graph.
\end{enumerate*}

In the following sections,
we present the problem formulation for heterogeneous graph summarization based on the MDL,
and how to design an optimization objective for the summary.
We describe how to encode the connectivity of each structure,
hierarchical labels, and errors.
Then, we describe how to generate a summary of the graph based on the formulation.
We describe subgraph generation,
subgraph identification based on the encoding cost of structures,
structure segmentation for label consistent structures by considering hierarchical labels,
and model construction approaches.
Table~\ref{tab:notation} shows the symbols used. % in this paper.

\begin{table} [t!]
	\centering
	\caption{Symbol description.}
	\label{tab:notation}
	\footnotesize
%	\resizebox{\columnwidth}{!}{%
		\begin{tabular}{cl}
			\toprule
			\textbf{Symbol} & \textbf{Description} \\
			\midrule
			$G=(\mathcal{V}, \mathcal{E})$ & Graph with node set $\mathcal{V}$ and edge set $\mathcal{E}$\\
%			$\mathcal{V}$,$|\mathcal{V}|$ & node-set, \# of nodes of $G$ resp.\\
%			$\mathcal{E}$,$|\mathcal{E}|$ & edge-set, \# of edges of $G$ resp.\\
			$\Omega$ & Set of structure types\\
			$st,ch$ & Star \& chain resp. \\
			$fc,nc$ & Full \& near clique resp.\\
			$bc,nb$ & Full \& near bipartite core resp.\\
%			$ch$ & Chain\\
			$M$ & Model\\
			$\mathcal{M}$ & Set of all possible models\\
			$\mathbf{A}$ & Adjacency matrix of $G$\\
			$s$ & Structure\\
			$area(s)$ & Edges of $G$ included in a structure $s$\\
%			$\mathcal{C}$ & set of all possible structures\\
%			$\mathcal{C}_x$ & set of all possible structures for type $x\in\Omega$\\
			$\mathbf{M}$ & Approximate adjacency matrix of $\mathbf{A}$ induced by $M$\\
			$\mathbf{E}$ & Error matrix, $\mathbf{E} = \mathbf{M} \oplus \mathbf{A}$\\
			$\oplus$ & Exclusive OR\\
			$L(G,M)$ & \# of bits to encode $G$ using $M$ \\
			$L(M)$ & \# of bits to encode model $M$ \\
			$L_t(s)$ & \# of bits to encode a structure $s$ \\
			$L_a(s)$ & \# of bits to encode node labels in a structure $s$ \\	
			$|s|$ & \# of nodes in a structure $s$\\
			$h$ & \# of levels \\
			$\mathbf{E}^+,\mathbf{E}^-$ & Edges that were over-modeled and under-modeled resp. \\
			$L(\mathbf{E}^+),L(\mathbf{E}^-)$ & Error encoding cost for $\mathbf{E}^+$ and $\mathbf{E}^-$ resp. \\
			$L(\mathbf{E}^a)$ & Encoding cost for labeling error \\
%			$|ne|$ & \# of nodes in labeling error node set \\
%			$|nel_{i,j}|$ & \# of nodes in \textit{l.e.n.s} having label $j$ in the $i$-th level\\
			\bottomrule
		\end{tabular}
%	}
\end{table}

%Given a heterogeneous graph $G=(\mathcal{V},\mathcal{E})$ with hierarchical node labels,
%we aim to construct a model (summary) $M$ using MDL principle.
%The model $M$ contains structures having types from a set $\Omega$ of structure types consisting of 6 most common structures found in real-world graphs~\cite{KoutraKVF14}:
%%stars ($st$),  \textit{full} and \textit{near} cliques ($fc$ and $nc$), \textit{full} and \textit{near} bipartite cores ($bc$ and $nb$),  and chains ($ch$).
%stars,  \textit{full} and \textit{near} cliques, \textit{full} and \textit{near} bipartite cores,  and chains.

%We elaborate on these structure types in Section~\ref{sec:mdl:connectivity}.
%Concisely, we have $\Omega = \{st,fc,nc,bc$ $,nb,ch\}$.

%where $\mathcal{M}$ is the set of all possible models, the best model $M \in \mathcal{M}$ for given data $D$ is the one that minimizes $L(M)+L(D|M)$, where $L(M)$ is the length in bits of $M$, and $L(D|M)$ is the length in bits of $D$ when encoded using $M$.

%We describe our problem formulation using MDL in Section ~\ref{sec:mdl_problem} and how to encode it in bits in Section ~\ref{sec:encoding_model}.

%To apply MDL for graph summarization using $\Omega$, we define what our possible models $\mathcal{M}$ are, how a model $M\in\mathcal{M}$ describes the given graph $G$, and how to encode it in bits.

\subsection{MDL for Heterogeneous Graph Summarization with Hierarchical Labels}
\label{sec:mdl_problem}

Minimum Description Length (MDL) ~\cite{RISSANEN1978465} is a model selection method where the best model $\hat{M}$ is the one that gives the best lossless compression:
%\vspace{-1mm}
	\begin{align*}
		\hat{M} = \argmin_{M \in \mathcal{M}}{(L(M) + L(D|M))}
	\end{align*}
%	\vspace{-1mm}
$M$ is a model, $\mathcal{M}$ is the set of all possible models, $D$ is the given data, $L(M)$ is the length in bits of $M$, and $L(D|M)$ is the length in bits of $D$ when encoded using $M$.
%MDL principle comes from the principle of Occam's razor, where the simplest model is considered the best one, and thus a model $M$ that minimizes $L(M) + L(D|M)$ is a "good" summary.

Our target model $M \in \mathcal{M}$ is an ordered list of graph structures each having a type from $\Omega$~\cite{KoutraKVF14}.
Each structure $s \in M$ corresponds to a certain portion of the adjacency matrix $\mathbf{A}$ describing its node's labels and interconnectivity;
there may be overlapped nodes between different structures, but there is no edge overlap.
Each node has a hierarchical label (e.g., Table~\ref{tab:label}).

%Connectivity of a structure is represented as $area(s,M,\mathbf{A})$, describing the included edges.
%We omit $M$ and $\mathbf{A}$, writing only $area(s)$ when the context is clear.

%Let $\mathcal{C}_x$ be the set of all possible structures for structure type $x \in \Omega$.
%For instance, $\mathcal{C}_{st}$ is the set of all possible stars.
%Our set of models $\mathcal{M}$ consists of all possible permutations of $\mathcal{C}_x$, where $\mathcal{C}_x$ is the set of all possible structures%
%Our model family $\mathcal{M}$ consists of all possible combinations of subsets of $\mathcal{C} = \cup_x \mathcal{C}_x$.
%In other words, the model family $\mathcal{M}$ includes all possible models $M$ of all structure types in a defined structure type set $\Omega$.

Following the MDL principle, our method for transmitting the adjacency matrix $\mathbf{A}$ is the following.
We 1) transmit model $M$,
2) induce the approximate adjacency matrix $\mathbf{M}$ of $\mathbf{A}$ by filling out the connectivity of each structure in $M$,
3) transmit the error $\mathbf{E}$ which is derived from taking the exclusive OR between $\mathbf{M}$ and $\mathbf{A}$ as MDL requires lossless encoding,
and
4) transmit labeling error $\mathbf{E}^a$ which is the label information for nodes not included in model $M$.
%Note that the reconstruction of the adjacency matrix $\mathbf{A}$ can be achieved in a lossless manner when the recipient knows $M$ and $\mathbf{E}$.
%We elaborate on label error $\mathbf{E}^a$ in Section ~\ref{sec:error}.
%
%Using MDL,
We opt for a model $M$ that minimizes the encoding length of model $M$, error $\mathbf{E}$, and labeling error $\mathbf{E}^a$.
Our problem definition is as follows.

%\vspace{-1mm}
\begin{problem} [Minimum Hierarchical Graph Description]
	$\newline$
	Given a heterogeneous graph $G$ with hierarchical labels on its nodes,
    its adjacency matrix $\mathbf{A}$,
    and set of structure types $\Omega$,
    find model $M$ that minimizes the total encoding length
	\begin{align*}
		L(G,M) = L(M) + L(\mathbf{E}) + L(\mathbf{E}^a)
	\end{align*}
	where $\mathbf{E}$ is the error matrix derived by $\mathbf{E} = \mathbf{M} \oplus \mathbf{A}$,
$\mathbf{M}$ is the approximate adjacency matrix of $\mathbf{A}$ induced by $M$, and
$\mathbf{E}^a$ is the labeling error.
\end{problem}
%\vspace{-1mm}
%

\subsection{Encoding the Model}
\label{sec:encoding_model}
Full encoding length of a model $M \in \mathcal{M}$ is the following:
\begin{align}
\label{eq:fullencoding}
	\begin{split}
	L(M) \;=\; &L_\mathbb{N}(|M|+1) + \log{|M|+|\Omega|-1 \choose |\Omega|-1}
	\\ &+ \sum_{s\in M}{(-\log(P(x(s)|M))+L_t(s)+L_a(s))}
	\end{split}
\end{align}

We encode
1) the total number of structures in $M$ using $L_\mathbb{N}$, Rissanen's optimal encoding~\cite{rissanen1983universal} for integers greater than $0$,
2) the number of structures for each type using an index over a weak number composition for each structure type $x \in \Omega$ in model $M$,
and
3) the structure type $x(s)$ for each structure $s \in M$ using optimal prefix code,
4) its connectivity $L_t(s)$,
and
5) the hierarchical labels $L_a(s)$ of nodes in it.
The detailed encodings for connectivity and hierarchical labels are described in the following two sections, respectively.

\subsection{Encoding Connectivity}
\label{sec:mdl:connectivity}

We describe how to encode the connectivity of each structure $s$ and derive its cost $L_t(s)$.

\noindent\textbf{Stars.} A star has a distinct characteristic of having a single "hub" node surrounded by 2 or more "spoke" nodes.
We compute $L_t(st)$ of a star $st$ as follows, where $|st|$ is the number of nodes inside the star and $|\mathcal{V}|$ is the number of nodes inside the given graph $G$:
\begin{align*}
	L_t (st) \;=\; L_\mathbb{N}(|st|-1) + \log |\mathcal{V}|+\log {|\mathcal{V}|-1 \choose |st|-1}
\end{align*}
We encode 1) the number of spokes in the star,
2) the index of the hub node over all $|\mathcal{V}|$ nodes,
and
3) which nodes are included in the star's spokes (spoke id).

\noindent\textbf{Cliques.}
A full clique is a subset of vertices where every two distinct vertices are adjacent.
We compute $L_t(fc)$ of a full clique $fc$ as follows, where $|fc|$ is the number of nodes inside the full clique:
\begin{align*}
	L_t(fc)\; =\; L_\mathbb{N}(|fc|) + \log {|\mathcal{V}| \choose |fc|}
\end{align*}
We encode 1) the number of nodes in the full clique, and 2) the index of a permutation to select $|fc|$ nodes out of $|\mathcal{V}|$ nodes.
%
%Note that the $fc$ that is considered might not actually be a full clique as $M$ is an approximation of a given graph $G$.
%From a encoding cost perspective, it might be advantageous to encode the given structure as a full clique and separately encode its missing or added edges as error.
%For the following case, each falsely represented edges will be encoded as errors which will be elaborated in Section ~\ref{sec:error}.

\noindent\textbf{Near cliques.}
A near clique  has several missing edges from a full clique.
We compute $L_t(nc)$ of a near clique $nc$ as follows:
\begin{align*}
	\begin{split}
 	L_t (nc) \;=\; & L_\mathbb{N}(|nc|) + \log {|\mathcal{V}| \choose |nc|}
 	 + \log(|area(nc)|)
 	+ ||nc||\epsilon_1 + ||nc||'\epsilon_0
	\end{split}
\end{align*}
where $|nc|$ represents number of nodes inside $nc$, and $|area(nc)|$ is the number of edges in $nc$.
We use $||nc||$ and $||nc||'$, respectively, for the number of observable and non-observable edges in $nc$.
$\epsilon_1$ and $\epsilon_0$ represent the lengths of the optimal prefix code for observable and non-observable edges, respectively.
\begin{equation*}
\begin{aligned}[c]
	\epsilon_1 = -\log(\frac{||nc||}{||nc||+||nc||'})\;
\end{aligned}\;\;
\begin{aligned}[c]
	\;\epsilon_0 = -\log(\frac{||nc||'}{||nc||+||nc||'})
\end{aligned}
\end{equation*}

\noindent We encode 1) the number of nodes in the near clique, 2) the index of a permutation to select $|nc|$ nodes out of $|\mathcal{V}|$ nodes,
%Compared to a full clique we need to encode the edges in a near clique, as some edges are missing from a full clique.
3) the number of edges,
4) the observable edges using the optimal prefix code, and
5) the non-observable edges.
%
%Intuitively, as a clique gets denser, the cheaper its encoding cost becomes.
%Note that the encoding for $nc$ is exact and no edges are added to the error $\mathbf{E}$.

\noindent\textbf{Bipartite cores.}
A bipartite core consists of two sets of nodes $A$ and $B$ where edges exist only between the sets and not within them.
A full bipartite core is a fully connected bipartite core. % where the two sets of nodes are fully connected.
We compute $L_t(fb)$ of a full bipartite core $fb$ as follows. %
%$|A|$ and $|B|$ are the number of nodes inside node sets $A$ and $B$ respectively:
\begin{align*}
	L_t(fb)\; =\; L_\mathbb{N}(|A|) +L_\mathbb{N}(|B|) + \log {|\mathcal{V}| \choose |A|}+\log {|\mathcal{V}| \choose |B|}
\end{align*}
where we encode 1) the number of nodes in node sets $A$ and $B$, and 2) ids of nodes in $A$ and $B$.

\noindent\textbf{Near bipartite cores.}
A near bipartite core has several missing edges from a full bipartite core.
% Similar to the near clique, a near bipartite core are also interesting if they protrude from the background distribution.
We compute $L_t(nb)$ of a near bipartite core $nb$ as follows, similarly to a near clique:
\begin{align*}
	\begin{split}
 	L_t (nb) \;=\; & L_\mathbb{N}(|nb|) + \log {|\mathcal{V}| \choose |nb|}
 	 + \log(|area(nb)|)
 	+ ||nb||\epsilon_1 + ||nb||'\epsilon_0
	\end{split}
\end{align*}
%
%Note that the encoding is exact and no edges are added to the error $\mathbf{E}$ as in near cliques.

\noindent\textbf{Chains.}
A chain is a sequence of nodes where each node is connected to the previous and the next node of it.
%For instance, consider a node set ${1,2,3}$. In this case node $1$ is connected to node $2$ and node $2$ is connected to node $3$.
%When plotted on an adjacency matrix this results in a super-diagonal non-zero elements.
We compute $L_t(ch)$ of a chain $ch$ as follows, where $|ch|$ is the number of nodes in $ch$:
\begin{align*}
	L_t(ch)\; =\; L_\mathbb{N}(|ch|-1) +\sum_{i=1}^{|ch|}{\log {(|\mathcal{V}|-i+1)}}
\end{align*}
We encode 1) the number of nodes in the chain, and 2) their ordered connectivity.

\begin{figure}[t]
	\centering
	 \subfloat{\includegraphics[width=0.4\textwidth]{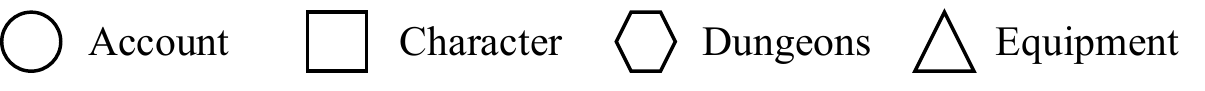}} \\
	 \vspace{-3mm}
	 \setcounter{subfigure}{0}
	\subfloat[\textit{Consistent} star] {\includegraphics[width=0.13\textwidth]{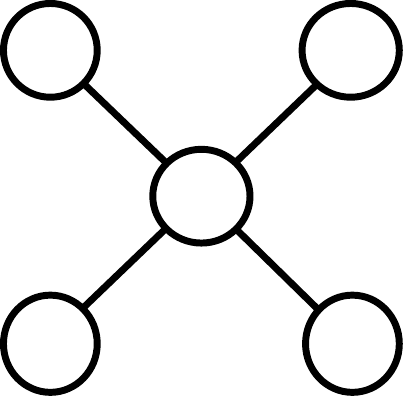}\label{fig:label_consistent}} \quad\quad
	\subfloat[\textit{Role-consistent} star] {\includegraphics[width=0.12\textwidth]{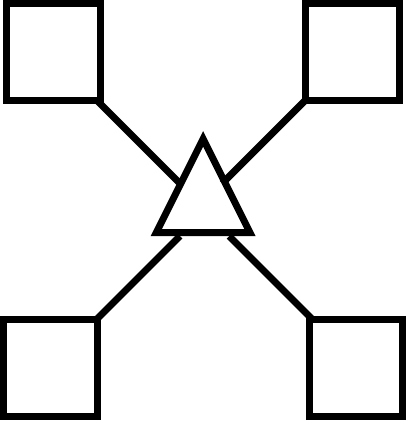}\label{fig:label_role_consistent}} \quad\quad
	\subfloat[Inconsistent star] {\includegraphics[width=0.13\textwidth]{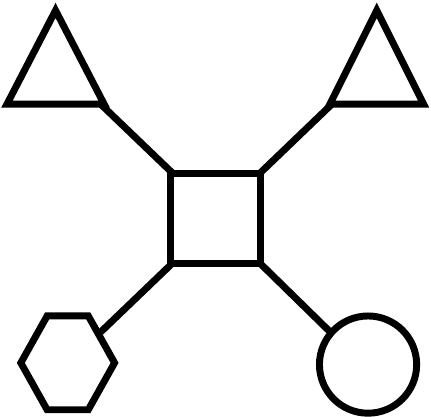}\label{fig:label_inconsistent}}
	\caption{Examples of label consistency in \method.}
\label{fig:label_consistency}
\end{figure}

\subsection{Encoding Hierarchical Labels}
\label{sec:hilabel}
We describe how to encode the hierarchical labels for the nodes in a structure $s$ and derive its cost $L_a(s)$.
%
%As discussed in Section ~\ref{sec:data},
Each node in our target graph has a hierarchical label:
e.g., a character node may have a three-level hierarchical label "character-tanker-blade master".
Furthermore, we assume that a different node may have a different depth of its label:
e.g., a dungeon node may have only two-levels of hierarchy, like "dungeon-advanced".
%
%Entities in most real-world data has properties and interact with other entities, this results in a heterogeneous graph.
%Furthermore, an entity could hold multiple properties at one.
%Hierarchical label includes multiple concepts inside a property and divides it into levels.
%For example, \textit{blade master} contains properties like \textit{character} and \textit{tanker} (see Table ~\ref{tab:label}), represented as one label: "character-tanker-blade master", where each level represents a partial concept of the whole property.
%
When transmitting a graph to the recipient, we assume that the recipient also knows how the hierarchical labels are structured.
We start by formalizing label encoding costs for node labels in level-$k$, and
describe a full label encoding cost $L_a(s)$ for hierarchical labels.

\textbf{Base encoding.}
In level-$k$, the base label encoding cost of a structure $s$ %$L^{base}_a(s,k)$
is given by
%$L^{base}_a(s,k)= \sum_{m=1}^{|s|}{\log(l^{b}_{k,m})}$,
\begin{align*}
	\begin{split}
	L^{base}_a(s,k)= \sum_{m=1}^{|s|}{\log(l^{b}_{k,m})}
%	& L^{base}_a(s,k)
%	\;=\; \sum_{m=1}^{|ns|}{\log(l^{b}_{k,m})}
	\end{split}
\end{align*}
where $|s|$ represents the number of nodes in the structure $s$.
$l^b_{k,m}$ is the number of unique \textit{sibling labels} for the level-$k$ label of the $m$-th node of the structure, where
\textit{sibling labels} of a label are those with the same upper level label (e.g. \textit{dealer}, \textit{tanker}, and \textit{buffer} are \textit{sibling labels} as they branch from \textit{character}).
For example, $l^b_{1,m}$, $l^b_{2,m}$, and $l^b_{3,m}$ are $4$, $3$ and $5$, respectively,
when the labels of the $m$-th node are \textit{character}-\textit{dealer}-\textit{destroyer} (see Table ~\ref{tab:label}).
%We encode each node's label in level-$k$ using the code $L^{base}_a(s,k) = \sum_{m=1}^{|ns|}{\log(l^{b}_{k,m})}$.

%, and $|nl_{i}|$ the number of nodes having label $i$.
%We encode 1) each label's count by an index over a weak number composition using $L^{w}_a(s) = \log{|ns| + l - 1 \choose l -1}$,
%and 2) each node's label using the code $L^{base}_a(s) = \sum_{m=1}^{|ns|}{\log(l)}$.
%Lastly, using binary representation of positive integers we encode other level's labels.

\textbf{Label consistency.}
We define the label encoding costs for two special cases: 1) \textit{consistent} structure, and 2) \textit{role-consistent} structure.
The two special cases allow us to reduce the label encoding cost.
A \textit{consistent} structure is the one containing only nodes with the same label, as shown in Fig.~\ref{fig:label_consistent}.
A \textit{role-consistent} structure is the one where nodes with the same \textit{role} have the same label, as shown in Fig.~\ref{fig:label_role_consistent}.
We define 4 roles in this work:
1) hub in a star structure,
2) spoke (neighbor of a star) in a star structure,
3) node belonging to the first set in a bipartite core,
and
4) node belonging to the second set in a bipartite core.
An inconsistent structure is the one that is neither consistent nor role-consistent, as shown in Fig.~\ref{fig:label_inconsistent}.

The label encoding cost %$L^{c}_a(s,k)$
for a \emph{consistent} structure in level-$k$ is given by
%$L^{c}_a(s,k) = \log(l^c_k)$,
\begin{align*}
	L^{c}_a(s,k) = \log(l^c_k)
\end{align*}
where $l^c_k$ of a consistent structure is the number of unique \textit{sibling labels} for the level-$k$ label.
For a consistent structure,
substituting $L^{c}_a(s,k)$ for $L^{base}_a(s, k)$ reduces the label encoding cost.
% replacing $L^{base}_a(s)$ with $L^{c}_a(s)$.
%Just by indexing over a weak number composition for the number of unique labels in level-1 $|l_1|$ the label type is implicitly defined.
%Furthermore, a star structure's spoke labels are all identical but labels of hub and spokes are different, or a (\textit{full} or \textit{near}) bipartite core's node-set $A$ and $B$ each have different but consistent labels throughout each set.
%We consider those structures as \textit{role-consistent} structures and

Next, the label encoding cost %$L^{rc}_a(s,k)$
for a \textit{role-consistent} structure in level-$k$ is given by
%$L^{rc}_a(s,k) = \log(l^{rc}_{k,1}) + \log(l^{rc}_{k,2}-1)$
\begin{align*}
	L^{rc}_a(s,k) = \log(l^{rc}_{k,1}) + \log(l^{rc}_{k,2}-1)
\end{align*}
where $l^{rc}_{k,1}$ and $l^{rc}_{k,2}$ of a \textit{role-consistent} structure are the number of unique \textit{sibling labels} for the level-$k$ labels of the two roles that a role-consistent structure can have, respectively.
%hub's (or node set $A$'s) label.
%Then, encoding the spoke's (or node set $B$'s) label.
%We reduce the label encoding cost $L_a(s)$ using $L^{rc}_a(s)$ instead of $L^{base}_a(s)$ when a structure is \textit{role-consistent}.
For a role-consistent structure,
substituting $L^{rc}_a(s,k)$ for $L_a(s,k)$ reduces the label encoding cost.
%First analogous to the \textit{consistent} case, number of unique labels in level-1 $|l_1|$ is encode using a weak number composition.

\textbf{Hierarchical label.}
%Based on the above label encoding cost, we extend the encoding cost to consider hierarchical labels.
%\blue{
%For hierarchical labels,
%we define the label encoding costs $L^{c}_a(s,k)$, $L^{rc}_a(s,k)$, and $L^{base}_a(s,k)$ of \textit{consistent}, \textit{role-consistent}, and \textit{inconsistent} structures in level-$k$:
%\begin{align*}
%	& L^{c}_a(s,k) \;=\; \log(l^c_k) \\
%	& L^{rc}_a(s,k) \;=\; \log(l^{rc}_{k,1}) + \log(l^{rc}_{k,2}-1) \\
%	& L^{base}_a(s,k) \;=\; \sum_{m=1}^{|ns|}{\log(l^{b}_{k,m})}
%\end{align*}
%$l^{rc}_{k,1}$ and $l^{rc}_{k,2}$ of a \textit{role-consistent} structure are the unique number of \textit{sibling labels} for the level-$k$ label of two roles.
%}
%
We define the full encoding cost $L_a(s)$ for hierarchical labels as follows:
\begin{align*}
	\begin{split}
	&L_a(s) \;=\; \log{|s| + l_1 - 1 \choose l_1 - 1}
	+ \sum_{i=1}^{h_1}{L^{c}_a(s,i)} + \sum_{j=h_1+1}^{h_2}{L^{rc}_a(s,j)} \\
	&+ \Big( \sum_{k=h_2+1}^{h}{L^{base}_a(s,k)} \Big) + 2\log(h) %\\
%	&\;=\;
%	\log{|ns| + l_1 - 1 \choose l_1 - 1}
%	+ \sum_{i=1}^{h_1}{\log(l^c_i)}
%	+ \sum_{j=h_1+1}^{h_2}{\left(\log(l^{rc}_{j,1})+\log(l^{rc}_{j,2}-1)\right)} \\
%	&+ \sum_{k=h_2+1}^{h} {\sum_{m=1}^{|ns|}{\left( \log (l^b_{k,m})\right)}} + 2\log(h)
	\end{split}
\end{align*}
where $h_1$ and $h_2$ mark the level in which consistent and role-consistent structures end, respectively; $h_1$ and $h_2$ are set to $0$ when consistent and role-consistent cases do not exist, respectively.
$l_1$ is the number of unique labels in level-$1$ and
$h$ is the number of levels.
We encode
1) the number of each label at level-$1$ by using an index over a weak number composition,
2) \textit{consistent} cases' labels from level-$1$ to level-$h_1$,
3) \textit{role-consistent} cases' labels from level-$(h_1 + 1)$ to level-$h_2$,
4) remaining levels using the base encoding for inconsistent cases, and
5) $h_1$ and $h_2$.
%We consider the three cases when moving from high (e.g. level-1) to lower (e.g. level-3) levels in the the following order: \textit{consistent}, \textit{role-consistent}, and \textit{inconsistent}.
We consider the cases in the order of \textit{consistent}, \textit{role-consistent}, and \textit{inconsistent} cases when moving from higher (e.g. level-1) to lower (e.g. level-3) levels.

\subsection{Encoding Errors}
\label{sec:error}
Given the summary $M$ of $G$, % and, $\mathbf{M}$ is the approximation of $\mathbf{A}$ induced by $M$,
it is vital to encode edges that are under or over-modeled by the model $M$ which we refer to as errors.
Specifically, there are two types of errors to consider.
The first type of error transpires in connectivity; if $area(s)$ is not identical to the regarding patch in $\mathbf{A}$, we encode the relevant errors.
Another error type arises from labeling, where nodes that failed to be included in at least a structure lack label information.

\subsubsection{Encoding Connectivity Errors}
The encoding length $L(\mathbf{E})$ of connectivity error of a model %$M$
is the following:
\begin{align*}
	& L(\mathbf{E}) \;=\;L(\mathbf{E}^+)+L(\mathbf{E}^-)
%	& \left(\log(|\mathbf{E}^+|)+||\mathbf{E}^+||\epsilon_1+||\mathbf{E}^+||'\epsilon_0\right)
%	+ \left(\log(|\mathbf{E}^-|)+||\mathbf{E}^-||\epsilon_1+||\mathbf{E}^-||'\epsilon_0\right)
\end{align*}
$\mathbf{E}^+$ and $\mathbf{E}^-$ are edges that are over-modeled and under-modeled, respectively;
$L(\mathbf{E}^+)$ and $L(\mathbf{E}^-)$ are the error encoding costs for $\mathbf{E}^+$ and $\mathbf{E}^-$, respectively.
%
%$\mathbf{E}^+$ represents edges that are not included in $\mathbf{A}$, but the model $M$ added in order to reduce a structure's connectivity encoding cost.
We encode $\mathbf{E}^+$ and $\mathbf{E}^-$ separately as they are likely to have different distributions.
Note that we ignore the errors from near cliques and near bipartite cores as they have been encoded exactly.
$L(\mathbf{E}^+)$ and $L(\mathbf{E}^-)$ are defined as follows.
\begin{align*}
	L(\mathbf{E}^+) \;=\;\log(|\mathbf{E}^+|)+||\mathbf{E}^+||\epsilon_1+||\mathbf{E}^+||'\epsilon_0\\
	L(\mathbf{E}^-) \;=\;\log(|\mathbf{E}^-|)+||\mathbf{E}^-||\epsilon_1+||\mathbf{E}^-||'\epsilon_0
\end{align*}
We encode first the number of 1s in $\mathbf{E}^+$ or $\mathbf{E}^-$,
and the 1s and 0s using the optimal prefix code.

\subsubsection{Encoding Labeling Errors}
\label{sec:labelerror}

We discuss encoding the labeling errors % $L(\mathbf{E}^a)$.
%Labeling error
which occur for nodes that the model %$M$
fails to cover, thus lacking the label information.
Our idea is to include all considering nodes into one structure and apply base encoding for label encoding cost $L_a(s)$.
Labeling error encoding cost $L(\mathbf{E}^a)$ is the following:
\begin{align*}
	\begin{split}
	L(\mathbf{E}^a) \;=\; &\log{|en| + l_1 -1 \choose l_1 -1}
	 + \sum_{k=1}^{h} {{\sum_{m=1}^{|en|} \log (l^b_{k,m})}}
%	\sum_{i=1}^{l_1}{|nel_{1,i}| \times \log(\frac{|nel_{1,i}|}{|ne|})}
%	\\&+ \sum_{i=2}^{h}{\sum_{j=1}^{l_j}{|nel_{i,j}| \times \log(|nel_{i,j}|)}}
	\end{split}
\end{align*}
where $|en|$ is the number of nodes in labeling error node set $en$, $l_1$ is the number of unique labels in level-$1$, and $l^b_{k,m}$ is the number of unique \textit{sibling labels} for the level-$k$ label of the $m$-th node.

\begin{algorithm} [t]
	\small
	\caption{Graph Summarization with Hierarchical Labels}
	\label{alg:method}
	\begin{algorithmic} [1]
%		\footnotesize
%		\small
%		\algsetup{linenosize=\small}
		\renewcommand{\algorithmicrequire}{\textbf{Input:}}
		\renewcommand{\algorithmicensure}{\textbf{Output:}}
		    \REQUIRE A heterogeneous graph $G$ with hierarchical node labels \\
		    \ENSURE Model $M$\\
		\STATE {\textbf{Subgraph generation.} Given the graph $G$, generate subgraphs using a graph decomposition method.}
		\STATE {\textbf{Subgraph identification.} For each subgraph obtained from step 1, we pick the structure which minimizes the local encoding cost of the subgraph.}
		\STATE {\textbf{Structure segmentation.} For each candidate structure, we segment it if the segmentation reduces the total encoding cost by considering the trade-off between the label encoding cost and the structure encoding cost.}
		\STATE {\textbf{Model summary.} We construct a model $M$
using summary approaches \textsc{Vanilla}, \textsc{Top-k}, and \textsc{Benefit}.}
	\end{algorithmic}
\end{algorithm}

\section{Proposed Method: Summary Generation}

%\begin{figure}[t]
%	\centering
%	\includegraphics[width=0.4\textwidth]{FIG/Example.pdf}
%	\caption{Example of a comparison between a large inconsistent structure and two small segmented structures.}
%	\label{fig:example}
%\end{figure}

We present \method (Graph Summarization with Hierarchical Labels), our proposed summarization algorithm for heterogeneous graphs with hierarchical node labels.
Algorithm~\ref{alg:method} describes \method.
%We describe the details of \method in the following sections.
%\subsection{Overview}
%\label{subsec:overview}
%
%There are several challenges to summarize heterogeneous graphs.
%\begin{itemize}
%	\item \textbf{Design structures with hierarchical labels.} How can we find meaningful strutctures for a subgraph with node labels?
%	\item \textbf{Define the encoding cost for hierarchical labels.} To summarize heterogeneous graphs using MDL principle, how can we design the encoding cost for hierarchical labels?
%\end{itemize}
%
%We suggest the main ideas to address the above challenges.
%
%\begin{itemize}
%	\item \textbf{Considering the role of nodes in a structure} provides meaningful structures, and reduce the encoding cost for labels.
%	\item \textbf{Considering a distribution of labels} reduces the encoding cost for the graph.
%\end{itemize}

%\begin{figure*}[t]
%        \centering
%        \subfloat[Example for star] {\includegraphics[width=0.3\textwidth]{FIG/Example.pdf}\label{fig:example_segement_star}}  \quad \tikz{\draw[-,black, densely dashed, thick](0,-0.5) -- (0,1.2);} \quad
%         \subfloat[Example for bipartite core]  {\includegraphics[width=0.63\textwidth]{FIG/Example2.pdf}\label{fig:example_segement_bc}}
%        \caption{Comparison between a large inconsistent structure and small segmented structures.}
%        \label{fig:example_segement}
%\end{figure*}

\subsection{Subgraph Generation}
\label{subsec:sub_generation}

We first decompose a given large graph into several subgraphs using a graph decomposition algorithm.
We leverage SlashBurn~\cite{KangF11} whose complexity is $O(T(m+n\log n))$ to generate subgraphs, where $m$ is the number of edges, $n$ is the number of nodes, and $T$ is the number of SlashBurn iterations.
SlashBurn has been used for many graph tasks~\cite{ShinJSK15,journals/tods/JungSSK16,JungPSK17};
\cite{KoutraKVF14,ShahKZGF15} demonstrate that SlashBurn successfully decomposes real-world graphs for graph summarization.
However, using other %graph decomposition
algorithms
like SUBDUE~\cite{CookH94} and BIGCLAM~\cite{YangL13} is possible.

\subsection{Subgraph Identification}
\label{subsec:subgraph_identification}

For each generated subgraph,
% (from Section~\ref{subsec:sub_generation}),
we find an appropriate structure in $\Omega$ to minimize the encoding cost. % for each subgraph.
For each subgraph, we 1) compute the encoding cost for all structure types, 2) compare the encoding costs, and 3) encode the subgraph as the structure type with the minimum encoding cost.
We assume that each subgraph corresponds to one structure among the six ones.

% Describe the local encoding cost.
We determine the role of each node in a given subgraph to encode the subgraph as one of the structure types.
For example, which node is the hub of a star?
Which nodes are included in set A or set B of a bipartite core?
How to determine the order of nodes in a chain?
We determine the role of each node %for each structure
in the following ways. %as follows.

\noindent\textbf{Stars.}
We encode the highest-degree node of the subgraph as the hub, and the remaining nodes as the spokes.

\noindent\textbf{Cliques.}
All nodes have the same structural role in a clique. %so that we simply encode a subgraph as a clique.

\noindent\textbf{Bipartite cores.}
We determine which nodes are included in set A or set B using binary
classification for bipartite cores~\cite{KoutraKVF14,ShahKZGF15}.
We add the highest degree nodes to set A, and its neighbors to set B.
Then, we perform Fast Belief Propagation (FaBP)~\cite{FaBP} to propagate classes and add nodes to set A or B.
%We assume that connected nodes have different classes and each class has several labels (e.g., account and equipment labels in set A).

\noindent\textbf{Chains.}
We adopt a heuristic approach~\cite{KoutraKVF14,ShahKZGF15} to encode a subgraph as a chain.
We
1) randomly pick a node and find a farthest node using breadth first search (BFS),
2) set the discovered node as starting node $n_{s}$,
3) find a farthest node from $n_{s}$ and set it as end node $n_{e}$, and
4) designate the shortest path from $n_{s}$ to $n_{e}$ as the initial chain.
Additionally, we extend the initial chain.
We construct an induced subgraph by removing all nodes included in the initial chain except for $n_{e}$.
We extend the chain by concatenating the shortest path from $n_{e}$ to a farthest node in the induced subgraph.
We repeat the aforementioned process for $n_{s}$.

% We extend the chain by concatenating the constructed path to the end node.
After obtaining the structures from a subgraph,
we compute and compare the encoding costs of them.
%Nodes and edges not included in the subgraph have minimal influence toward the subgraph identification, and thus we compute the local encoding cost by using only the given subgraph.
Given a subgraph $G'$, we compute the local cost $L(M_s) + L(\mathbf{E}_{M_s})$~\cite{KoutraKVF14,ShahKZGF15} to encode it as a structure $s$,
where $M_s$ is the model consisting only of the subgraph $G'$ encoded as $s$,
and
$\mathbf{E}_{M_s}$ is the error matrix derived by $\mathbf{E}_{M_s} = \mathbf{M}_s \oplus \mathbf{A}_{G'}$.
$\mathbf{A}_{G'}$ is the adjacency matrix of the subgraph $G'$ and
$\mathbf{M}_s$ is the approximate adjacency matrix of $\mathbf{A}_{G'}$ induced by $M_s$.
Note that the labeling errors are ignored since nodes in a structure are connected and thus there are no uncovered nodes inside the structure.
We compute $L(M_s)$ by adding $L_t(s)$ and $L_a(s)$ (e.g., $L_t(st) + L_a(st)$ for $st$).
The remaining terms except for $L_t(s)$ and $L_a(s)$ in Equation~\eqref{eq:fullencoding} are negligible, since $M_s$ consists only of the structure $s$.
$L(\mathbf{E}_{M_s})$ is equal to $L(E^{+}_{M_s})+L(E^{-}_{M_s})$;
$L(E^{+}_{M_s})$ and $L(E^{-}_{M_s})$ are the error encoding costs for $E^{+}_{M_s}$ and $E^{-}_{M_s}$, respectively,
which are edges being over-modeled and under-modeled in the given subgraph $G'$, respectively.

We add the best structure $s$ minimizing the local encoding cost $L(M_s) + L(\mathbf{E}_{M_s})$
to the set $\mathcal{C}$ of structures.

\begin{figure}[t]
        \centering
	\includegraphics[width=0.45\textwidth]{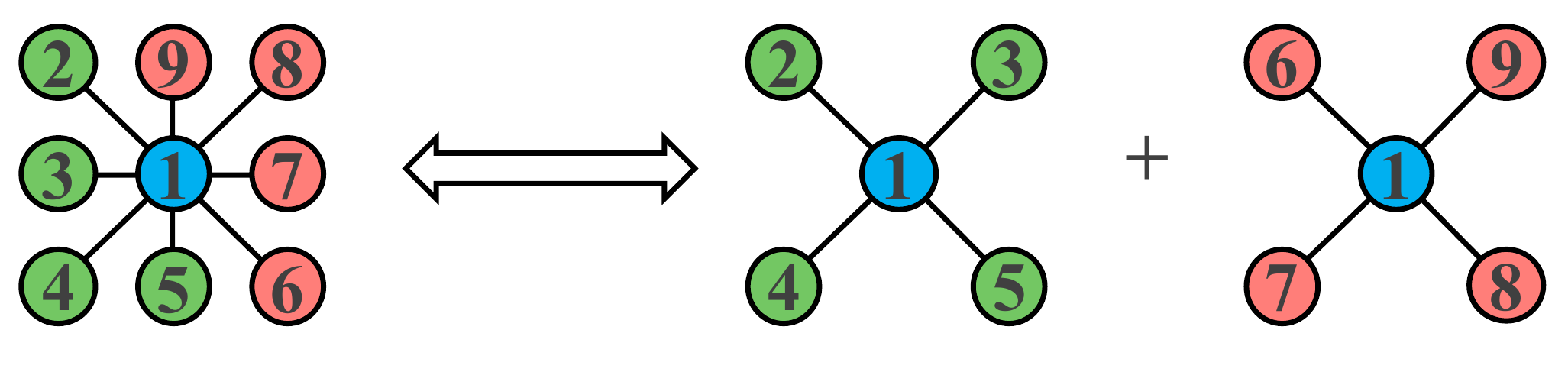}
        \caption{%\blue{Example for star.
\method chooses small segmented star structures in the right rather than a large inconsistent star in the left since the right ones minimize the encoding cost.
        }
        \label{fig:example_segement_star}
\end{figure}

\begin{figure*}[t]
        \centering
		\includegraphics[width=0.95\textwidth]{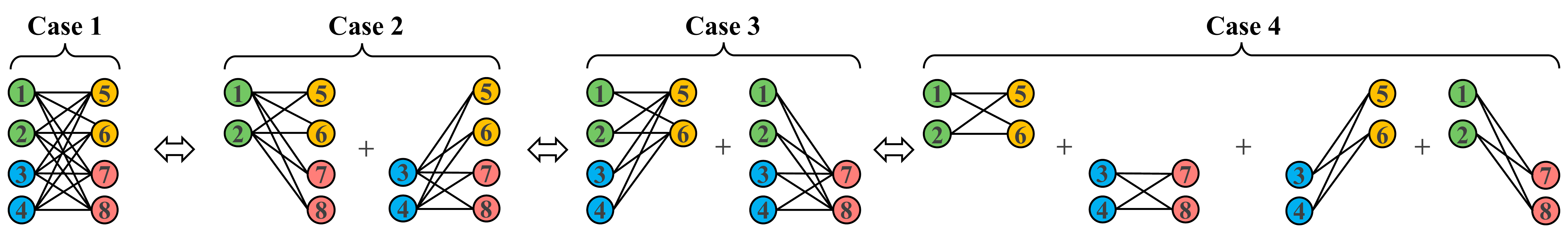}
        \caption{
\method chooses the best structure with the minimum encoding cost among the 4 possible cases of segmentation in a bipartite core.
        }
        \label{fig:example_segement_bc}
\end{figure*}

\subsection{Structure Segmentation}
\label{subsec:structure_seg}

We describe the segmentation of a structure considering hierarchical labels of nodes.
The objective is to further reduce the label encoding cost for a structure considering the consistency of labels.
Fig.~\ref{fig:example_segement_star} shows an example of comparison between a large \textit{inconsistent} structure and small \textit{consistent} structures.
The inconsistent star graph in the left with 9 nodes has a low structure encoding cost, but a high label encoding cost.
In contrast, the two decomposed star structures in the right have higher total structure encoding costs, but a lower total label encoding costs than the left case.
We compare the total encoding costs before and after segmentation,
and determine whether to segment or not by selecting the option leading to the minimum cost.
We describe how to segment each structure, and how to extend the segmentation idea to hierarchical labels in the following two subsection, respectively.
% The decision to segment the large structure is determined by the total encoding cost incurred after and prior to segmentation
%For example, a large \textit{inconsistent} structure having various labels has a low structure encoding cost, but a high label encoding cost.
%Contrary to the large \textit{inconsistent} structure, multiple small \textit{consistent} structures consisting of nodes in the large \textit{inconsistent} structure have high structure encoding costs, but low label encoding costs.

\subsubsection{Label Segmentation}
\label{subsubsec:label_segmentation}

We describe the segmentation of each structure in a level.
We divide a structure into smaller structures of the same type \emph{only when} the smaller structures provide a lower encoding cost.
%We assume that all nodes in the given structure are included in one of the smaller structures which have the same structure type as the given structure.
%We observe that small segmented graphs maintain the structure of a large subgraph.
%Therefore, we assume that we encode small segmented graphs as the same structure of a large subgraph, and then compute encoding costs of small segmented graphs.

\noindent\textbf{Stars.}
We divide spokes of a star into two groups considering the consistency of labels, and construct two stars which have the same hub, but different spokes.
We assign spoke nodes of the majority label to the first group and the remaining spoke nodes to the second group.

\noindent\textbf{Bipartite cores.}
Bipartite cores have two sets of nodes.
For each set, we search for the majority label in the set and divide the nodes of the set into the nodes with the majority label and the rest.
As shown in Fig.~\ref{fig:example_segement_bc}, we consider $4$ cases depending on the segmentation for each set: 1) no segmentation, 2) segmenting only the first set, 3) segmenting only the second set, and 4) segmenting both sets.
We choose the option with the minimum encoding cost among the 4 possible cases.

\noindent\textbf{Cliques and chains.}
Contrary to star and bipartite core structures, all the nodes in cliques and chains have the same role.
We first find the majority label, and then divide the structure into nodes of the majority label and the rest.
We segment the given structure when the sum of the encoding costs of the two segmented structures is lower than that of the given structure.

%\begin{algorithm} [t]
%	\caption{Hierarchical Segmentation}
%	\label{alg:hierarchical_segmentation}
%	\begin{algorithmic} [1]
%%		\footnotesize
%		\small
%		\algsetup{linenosize=\small}
%
%		\renewcommand{\algorithmicrequire}{\textbf{Input:}}
%		\renewcommand{\algorithmicensure}{\textbf{Output:}}
%		    \REQUIRE a structure $s$\\
%		\STATE {$Q \leftarrow \{s\}$}
%		\WHILE{$Q \not= \emptyset$}
%		\STATE{aa}
%		\ENDWHILE
%% 		\REPEAT
%% 			\FOR {$n=1,...,N$}
%% 				\STATE{$\T{Y} \leftarrow \T{X}\times_1\mat{A}^{(1)T}\cdots\times_{n-1}\mat{A}^{(n-1)T}\times_{n+1}\mat{A}^{(n+1)T}\cdots\times_N\mat{A}^{(N)T}$} \label{line:ttmc}
%% 				\STATE{$\mat{A}^{(n)} \leftarrow \mat{R}_n$ leading left singular vectors of $\mat{Y}_{(n)}$}
%% 			\ENDFOR
%% 		\UNTIL{the maximum iteration is reached, or the error ceases to decrease;}\\
%% 		\STATE{$\T{G} \leftarrow \T{X}\times_1\mat{A}^{(1)T}\times_{2}\mat{A}^{(2)T}\cdots\times_N\mat{A}^{(N)T}$} \label{line:core}
%	\end{algorithmic}
%\end{algorithm}

\subsubsection{Hierarchical Consideration}
\label{subsubsec:hierarchy_consideration}

Given a structure with hierarchical labels, we segment the structure from higher to lower levels. % based on the procedure of label segmentation described in Section~\ref{subsubsec:label_segmentation}.
For each level, we perform label segmentation to determine whether the structure should be segmented or not. % by comparing the encoding costs.
%Label segmentation performs the following tasks:
Hierarchical label segmentation follows the following rules.

\begin{enumerate*}
\item If a structure is \textit{consistent} or \textit{role-consistent} at the current level, the structure is challenged for label segmentation at the next lower level.
\item If an \textit{inconsistent} structure is segmented and the current level is not the lowest %(i.e., level-$3$ in Blade \& Soul graph),
    the segmented structures are independently challenged for segmentation at the next lower level.
\item If an \textit{inconsistent} structure is not segmented, we terminate the procedure and insert the structure in the set $\mathcal{C}$ of the candidate structures.
\item At the lowest level, we compare the encoding costs and insert the structure or the segmented structures in the set $\mathcal{C}$ of the candidate structures.
\end{enumerate*}

\subsection{Model Summary}
\label{subsec:model_summary}

We construct the final model which summarizes the given heterogeneous graph by carefully selecting structures among the candidate structures $\mathcal{C}$.
Finding the best model exactly by exploring all possible subsets of candidate structures
is intractable.
Thus we propose the following heuristics to choose a subset of candidate structures in a scalable way.
\begin{itemize}[noitemsep,topsep=0pt]
	\item \textsc{Vanilla}: All candidate structures are included in our summary, thus $M$ is equals to the set $C$ of the candidate structures.
	\item \textsc{Top-k}: We pick the top-$k$ structures by sorting the local encoding gains of candidate structures in descending order.
A local encoding gain is the number of bits reduced by encoding a subgraph as a structure $x$.
	\item \textsc{Benefit}: We pick all structures whose local encoding gains are greater than zero.
\end{itemize}

\method uses the above approaches to finalize the model.
%The comparison of these heuristics is discussed in Section~\ref{subsec:quant_analysis}.

\begin{table}[t]
%\small
\centering
\caption{%(The lower the better.)
Comparison of structures and costs of \method and the original graph.
\textsc{\method-Vanilla} explains 99\% of 7,885,487 edges using 25,740 structures.
\textsc{\method-Top-100} uses only 100 structures to explain 56\% of edges.
\textsc{\method-Benefit} uses 4,382 structures to explain 95\% of edges.
}
\label{tab:resultcost}
\resizebox{0.99\columnwidth}{!}{%
\begin{tabular}{lrrrr}
\toprule
 		&       & \textbf{Description }  	&  \textbf{Relative}	  & \textbf{Unexplained }\\
\textbf{Method} 		& \textbf{\# of structures}  & \textbf{cost (bits)}  	& \textbf{cost} 	  & \textbf{edges }\\
\midrule
\textsc{Original} & $-$ & $106,444,727$   			& $100\%$             & $0\%$    \\
\textsc{\method-Vanilla} &  $25,740$	 	& $52,124,308$     			& $49\%$     & $1\%$    \\
\textsc{\method-Top-100} &  $100$	 	& $75,781,149$     			& $71\%$              & $56\%$    \\
%\textsc{\method-Top-10000} 	& $58,015,747$     			& $55\%$              & $9\%$    \\
\textsc{\method-Benefit} &	 $4,382$ 	& $52,237,349$     			& $49\%$              & $5\%$    \\ \bottomrule
\end{tabular}
}
\end{table}

\section{Experiment}

%\begin{table}[!t]
%	\caption{Description of Blade \& Soul graph.}
%	\centering
%	\label{tab:Description}
%		\resizebox{0.3\textwidth}{!}{
%	\begin{tabular}{lrrr}
%		\toprule
%		\textbf{Dataset} & \textbf{Nodes} & \textbf{Edges} \\
%		\midrule
%		 Blade \& Soul Graph & $249,455$ & $7,885,487$  \\
%		\bottomrule
%	\end{tabular}}
%\end{table}

We evaluate \method to answer the following questions:

\begin{itemize}[noitemsep,topsep=0pt]
	\item [\textbf{Q1}] \textbf{Graph summarization.} How well does \method summarize the Blade \& Soul graph?
	\item [\textbf{Q2}] \textbf{Discovery.} What are the discovery results of analyzing Blade \& Soul graph with \method?
	\item [\textbf{Q3}] \textbf{Finding similar users.} How can we find similar users using the summary by \method?
 	\item [\textbf{Q4}] \textbf{Scalability.} How well does \method scale up in terms of the number of edges?
	\end{itemize}

\begin{table*}[ht]
\centering
\caption{Summarization results of Blade \& Soul graph by \method.
%We use three approaches of \method for summarization.
%We count the number of structures for each approach.
$st$, $fc$, $nc$, $bc$, $nb$, and $ch$ indicate star, full clique, near clique, bipartite core, near bipartite core, and chain, respectively.}
\label{tab:summary}

\resizebox{0.35\textwidth}{!}{
\subtable[\textsc{\method-Vanilla}]{
\begin{tabular}{cccccc}
\toprule
st & fc & nc & bc & nb & ch \\
\midrule
$22,787$  & $21$ &  $-$   &  $-$  &  $2,467$  & $465$ \\
\bottomrule
\end{tabular}
}}
\resizebox{0.28\textwidth}{!}{
\subtable[\textsc{\method-Top-100}]{
\label{tab:top100}
\begin{tabular}{cccccc}
\toprule
st & fc & nc & bc & nb & ch \\
\midrule
$100$  & $-$ &  $-$   &  $-$  &  $-$  & $-$ \\
\bottomrule
\end{tabular}
}}
%\resizebox{0.225\textwidth}{!}{
%\subtable[Top-$10000$]{
%\begin{tabular}{cccccc}
%\toprule
%st & fc & nc & bc & nb & ch \\
%\midrule
%$9,505$  & $18$ &  $-$   &  $-$  &  $12$  & $465$ \\
%\bottomrule
%\end{tabular}
%}}
\resizebox{0.3\textwidth}{!}{
\subtable[\textsc{\method-Benefit}]{
\begin{tabular}{cccccc}
\toprule
st & fc & nc & bc & nb & ch \\
\midrule
$3,917$  & $-$ &  $-$   &  $-$  &  $-$  & $465$ \\
\bottomrule
\end{tabular}
}}
\end{table*}

We use the Blade \& Soul graph for experiments.
%\method is implemented on Python.
We run experiments on a workstation with an Intel Xeon E5-2630 v4 @ 2.2GHz CPU and 512GB memory.

%\textbf{Graph Generation.}
%As described in Section~\ref{subsec:sub_generation}, we use SlashBurn~\cite{KangF11} to generate subgraphs since it deals with large-scale non-cavemen graphs.
%Figure~\ref{fig:data_analysis2} shows that SlashBurn is suitable for our graph as the adjacency matrix reordered by SlashBurn seems an arrow pointing to the upper left while having shallow wings.

\subsection{Graph Summarization (Q1)}
\label{subsec:quant_analysis}

We evaluate the description cost and edge coverage of Blade \& Soul graph by \method.
Our baseline is \textsc{Original} model where
the original edges are encoded with $L(\mathbf{E}^-)$ and the labels are encoded with $L(\mathbf{E}^a)$.
%We compute the description costs for \method with four selection heuristics (\textsc{VANILLA}, \textsc{TOP-100}, \textsc{TOP-10000}, and \textsc{BENEFIT}).

Table~\ref{tab:resultcost} compares the number of structures, the description cost, and
the ratio of unexplained edges by \textsc{Original} and the three proposed models (\textsc{\method-Vanilla}, \textsc{\method-Top-100}, and \textsc{\method-Benefit}).
%Lower the relative cost, more edges were covered.
%We evaluate our models based on the results of compression because a concise model is a good summary from an MDL's standpoint.
\textsc{\method-Vanilla} summarizes the given graph with only $49\%$ of the bits w.r.t. \textsc{Original} and explains all but $1\%$ of the edges which are not modeled by $M$.
%
%\textsc{\method-Vanilla} identifies significant structures in the given graph by considering the hierarchical labels.
%
\textsc{\method-Top-100} explains $44\%$ of the edges and summarizes the given graph with $71\%$ of the bits w.r.t. \textsc{Original}, while containing only $100$ structures out of $25,740$ structures.
Each structure in \textsc{\method-Top-100} contains a larger number of nodes and edges compared to the structures not included.
Compared to \textsc{\method-Vanilla}, \textsc{\method-Top-100} has fewer number of structures but a larger cost.
\textsc{\method-Benefit} shows a reasonable balance between the description cost and the number of structures.
Compared to \textsc{\method-Vanilla},
it has a similar description cost, but contains $5.9\times$ smaller number of structures.

\begin{figure}[t]
	\centering
	\includegraphics[width=0.8\textwidth]{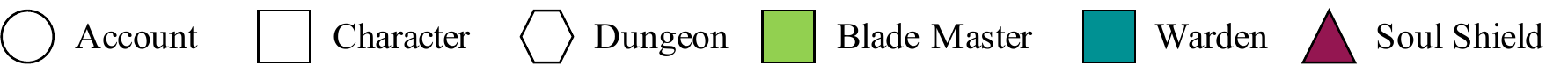}\\
	\vspace{-1mm}
	 	 \subfloat[$st$]{\includegraphics[width=0.2\textwidth]{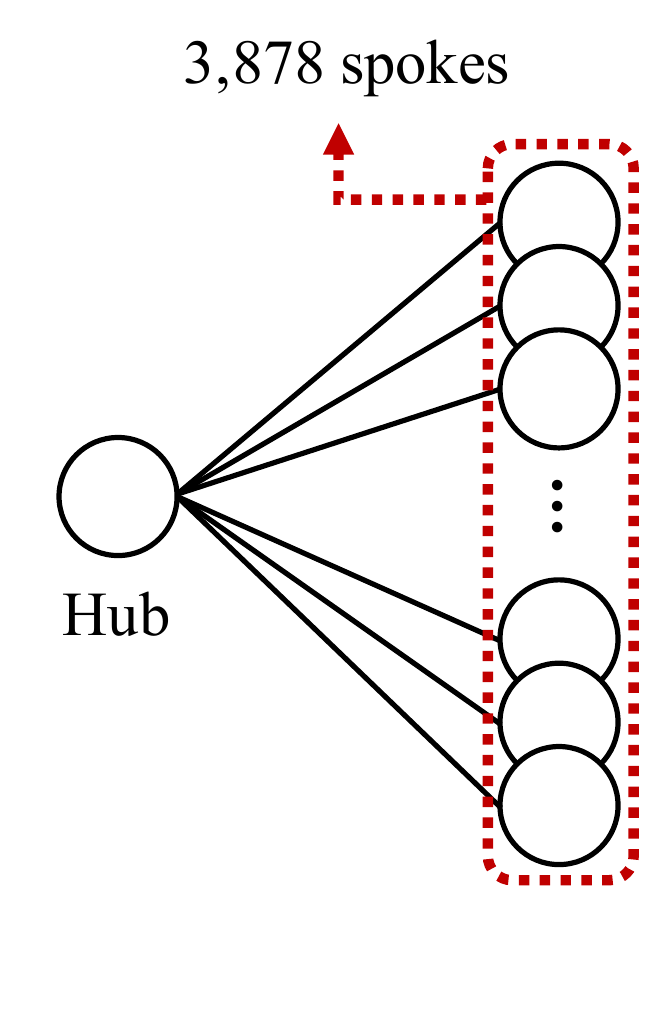}\label{fig:casest}}
	 	 \subfloat[$fc$]{\includegraphics[width=0.35\textwidth]{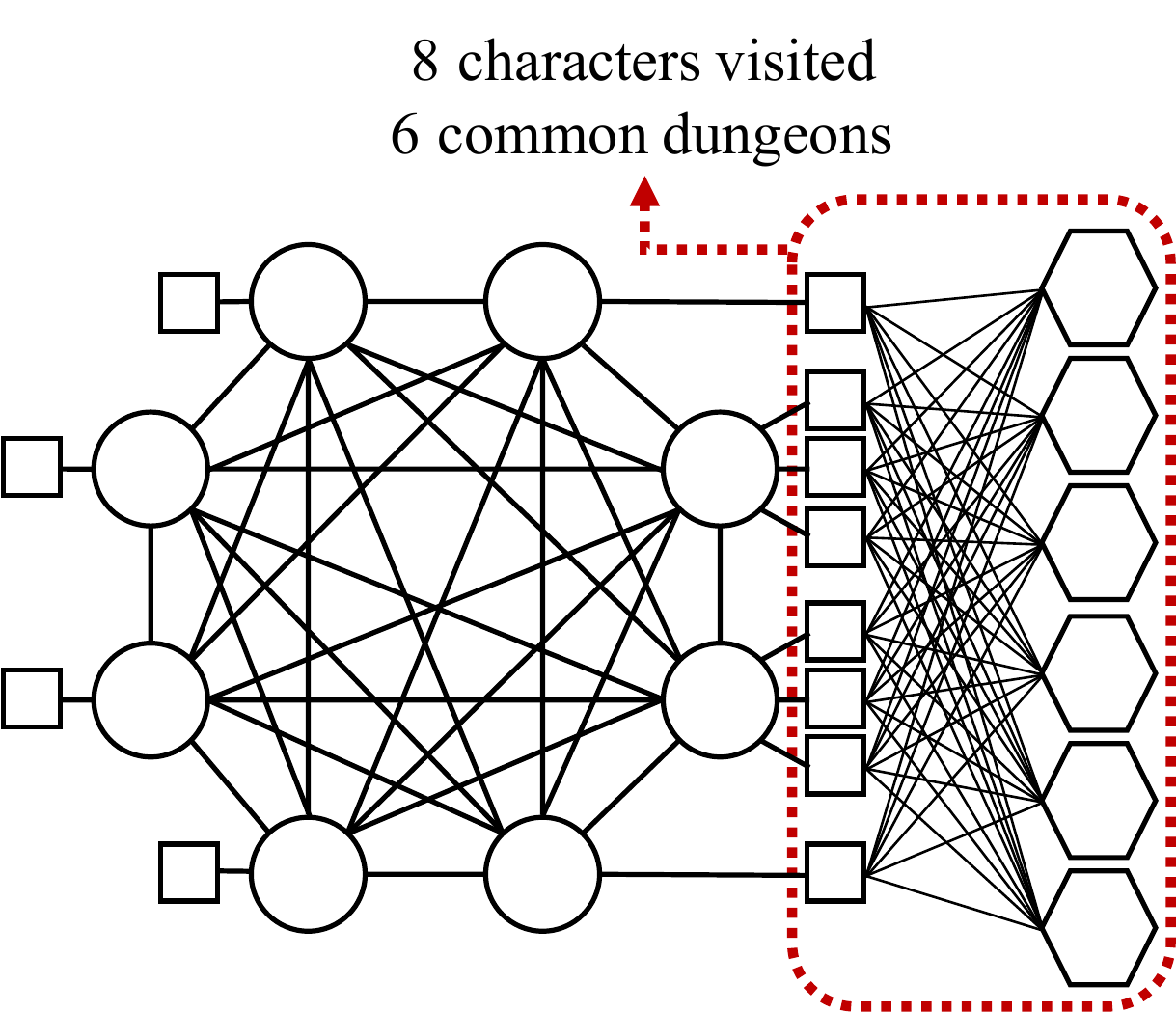}\label{fig:casefc}}
	 	 \subfloat[$nb$]{\includegraphics[width=0.2\textwidth]{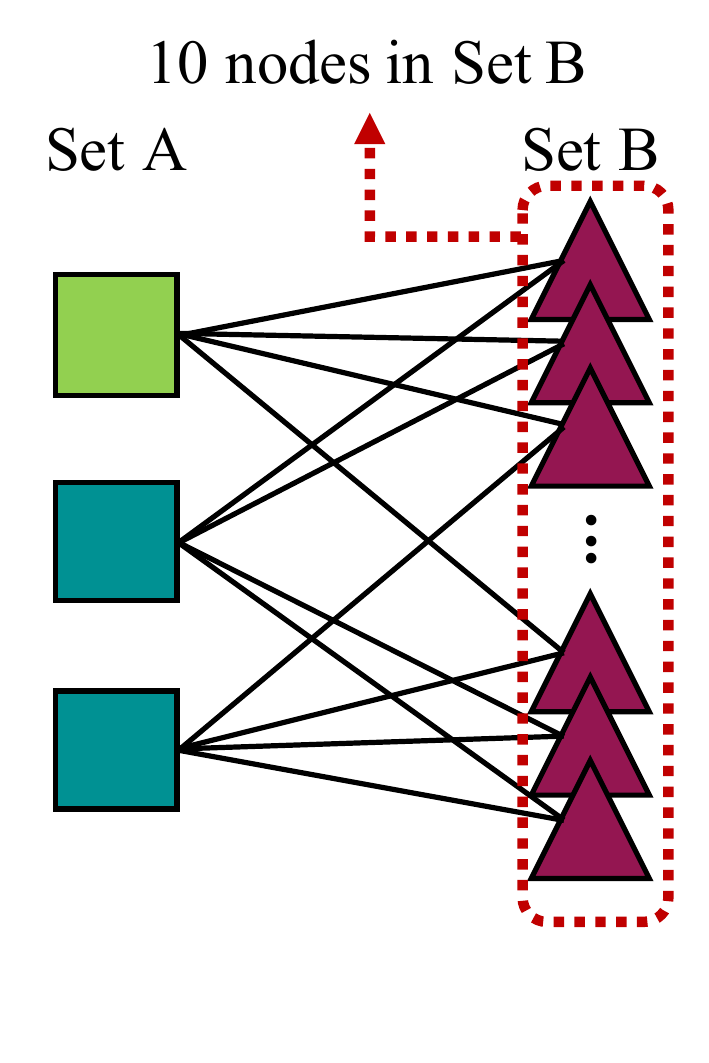}\label{fig:casenb}}
%	 	 \subfloat[{\color{red}Fake figure}]{\includegraphics[width=0.25\textwidth]{FIG/CASE_SEGMENTATION.pdf}\label{fig:case}}\\
	\caption{Meaningful structures from the summary.
	(a) $st$ (star) structure consisting only of accounts, where the hub node represents an influential \textit{account}.
%	(b) $nb$ with two accounts, have five Kung Fu Master characters, which are equipped with the same two equipment, which is interpreted as similar progress between the five characters.
%	(b) $fc$ structure with $8$ \textit{accounts} representing a small clan.
%
	(b) $fc$ (full clique) structure consisting of $8$ \textit{account} nodes constituting a small clan. %$4$ \textit{accounts} on the left possess inactive characters (never visited a dungeon) and
The $4$ \textit{accounts} on the right possess characters which visited $6$ common \textit{dungeons}. %, where the $fc$ represents a small clan.
%
%	(c) For the node (id: $249,444$) which has \textit{equipment}-\textit{soul} label, Zen Archer characters especially prefer this item to other characters although all the characters can equip this item.
	(c) $nb$ (near bipartite core) structure where a \textit{blade master} and two \textit{wardens} in set A (which are tankers) are equipped with several \textit{soul shields} from set B, representing similar preferences for the three \textit{characters} and similar properties for the $10$ \textit{soul shields}.
	}
	\label{fig:comprehension}
\end{figure}

% with the consistency of labels
%\textsc{Benefit} as it has a similar relative cost compared to \textsc{Vanilla}, yet having more unexplained edges.
%\textsc{Benefit} summarizes the given graph with fewer structures while acquiring a more succinct summary, as it has a similar relative cost compared to \textsc{Vanilla}, yet having more unexplained edges.

%While we allude to the description cost as a summarization technique, compression itself is not our goal, as from an MDL's standpoint; a good compression results in a good summary.
%For this reason we do not compare with standard matrix compression methods.
%\method has the goal of describing a graph with comprehensible structure, while some specialized algorithms exploit statistical correlations to save description cost.

\subsection{Discovery (Q2)}
\label{subsec:qual_analysis}

We present the discovery results of Blade \& Soul graph using \method.

\subsubsection{Major Structures}
How does \method summarize the Blade \& Soul graph, and which structures can we discover?
%We summarize the graph extracted from Blade \& Soul.
Table~\ref{tab:summary} shows the structures from our proposed methods
\textsc{\method-Vanilla}, \textsc{\method-Top-100}, and \textsc{\method-Benefit}.
Note that \textsc{\method-Vanilla} contains the most diverse structures including star, full cliques, near bipartite cores, and chains.
\textsc{\method-Top-100} contains only stars as they are the dominant structures when using SlashBurn for graph decomposition.
\textsc{\method-Benefit} contains mainly stars, but it also contains several chains; it means that there are clear chain-like structures in the decomposed subgraphs.

\subsubsection{Meaningful Structures}
What interpretations can we make out of the summary?
We report meaningful structures of Blade \& Soul graph discovered by \method.
%When observing structures sorted by local encoding gains in descending order,
%We observe that the superseding structures are $st$ structures, as in Table~\ref{tab:top100} \textsc{Top-100} only consists of 100 $st$ structures.

\noindent\textbf{Popular dungeons and equipment.}
We observe that the hub node labels of star structures in \textsc{\method-Top-100} mainly consist of popular \textit{dungeon} and \textit{equipment} nodes, surrounded only by \textit{character} nodes.
This is due to the nature of Blade \& Soul where \textit{characters} cooperate with other \textit{characters} visiting various \textit{dungeons}, while having various \textit{equipment}.
%We interpret that the hubs of these structures represent popular \textit{dungeons} or \textit{equipment}.

\noindent\textbf{Influential account.}
Fig.~\ref{fig:casest} shows a star structure consisting only of \textit{account} nodes.
The hub node is an influential \textit{account} connected to $3,878$ other \textit{accounts}.

\noindent\textbf{Small clan.}
%Fig.~\ref{fig:casefc} shows a full clique structure consisting only of \textit{account} nodes.
%This represents a small clan, where the \textit{accounts} frequently play with each other.
%
Fig.~\ref{fig:casefc} shows $8$ \textit{account} nodes forming a full clique structure.
%, where $4$ \textit{accounts} on the left possess inactive characters (never visited a dungeon) and
The $4$ \textit{accounts} on the right possess characters which visited $6$ common \textit{dungeons}.
This represents a small clan where the \textit{accounts} play together frequently.

\noindent\textbf{Near bipartite core of characters and equipment.}
Fig.~\ref{fig:casenb} shows a bipartite core;
a \textit{blade master} and two \textit{wardens} are \textit{tankers} in set A,
and they are equipped with several \textit{soul shields} in set B.
This indicates that all the three characters have the same preference for the equipment.

%a $nb$ with $5$ \textit{assassins} possessed by one of the two \textit{accounts}, while the two \textit{accounts} are not friends.
%These five \textit{assassins} are all equipped with the same \textit{ring}, meaning the two \textit{accounts} prefer the same \textit{equipment}.

%Figure ~\ref{fig:casenb} shows a $nb$ with $5$ \textit{assassins} possessed by one of the two \textit{accounts}, while the two \textit{accounts} are not friends.
%These five \textit{assassins} are all equipped with the same \textit{ring}, meaning the two \textit{accounts} prefer the same \textit{equipment}.

%Figure ~\ref{fig:casenb} shows a $bc$ with two \textit{accounts} having five \textit{Kung Fu Master characters}, which are equipped with the same two \textit{equipment}, which is interpreted as similar progress rate between the five \textit{characters}.
%Here we notice that this structure was encoded as a $bc$, but as there is only one unique edge between \textit{account} to \textit{character}, \method has decided to add the missing edges to the over-modeled edges $\mathbf{E}^+$, as it was beneficial to the local encoding cost of $bc$.

\noindent\textbf{Segmented star.}
We analyze structures segmented by considering hierarchical labels.
Fig.~\ref{fig:segmented} shows that
a large star structure whose hub is an equipment node with \textit{soul} label is divided into three smaller star structures.
The spokes of each star structure consist of:
a) \textit{zen archers},
b) \textit{dealers} except for \textit{zen archers}, and
c) all \textit{jobs} in \textit{tanker} and \textit{buffer}.
%
%Note that the label of the hub is \textit{equipment}-\textit{soul}.
This segmentation occurs since $37\%$ $(760/2041)$ of the spokes in the star structure are \textit{zen archers} (out of the $11$ total \textit{jobs}).
Although any \textit{jobs} can have this \textit{soul} equipment, \textit{zen archers} prefer this \textit{soul} equipment the most.
%This result indicates that characters of \textit{dealer} role prefer this equipment although any \textit{job} can equip it.
%Furthermore, \textit{zen archer} prefers this item to other \textit{jobs}.

\subsection{Finding Similar Users (Q3)}
\label{subsec:application}

How can we exploit the result of summarization for finding similar accounts in Blade \& Soul?
We measure account similarities as follows:

\begin{enumerate}[noitemsep,topsep=0pt]
	\item \textbf{Matrix construction.}
Given the found structures,
we first construct a \textit{node}-\textit{structure} matrix $\mathbf{B}$.
The $(i,s)$-th element of the matrix $\mathbf{B}$ is set to a value $\gamma^{\alpha_{i,s}}$
where
$\gamma$ is set to $0.7$ in this experiment and
$\alpha_{i,s}$ is the minimum length of the shortest paths between account $i$ and nodes in structure $s$.
If node $i$ is in structure $s$, $\alpha_{i,s}$ is set to $0$.
	\item \textbf{Randomized SVD for features.} We compute a randomized SVD for the matrix $\mathbf{B}$ to obtain features of accounts.
	\item \textbf{Cosine similarity.} We compute the cosine similarity
of the two account feature vectors.
\end{enumerate}
%Intuitively, two accounts are similar when
%1) there are many structures including both accounts, and
%2) there are many structures in the neighborhood of both accounts.

We pick a target account with id $0$, and find the two most similar accounts which have ids 3602 and 9.
We observe that they share common characters, equipment, and often friends.
Fig.~\ref{fig:similar_account3602} shows the relation of accounts 0 and 3602.
Note that they have the characters with the same jobs (destroyer and blade master).
The corresponding characters have common equipment and visited the same dungeons; e.g.,
the two destroyers have $21$ identical equipment, and visited $12$ identical dungeons.
Fig.~\ref{fig:similar_account9} shows the relation of accounts 0 and 9.
We observe the similar patterns of characters and their related equipment and dungeons.
They also share 18 common friends as well.

\subsection{Scalability (Q4)}
\label{subsec:scalability}

\begin{figure}[t]
	\centering
		\includegraphics[width=0.4\textwidth]{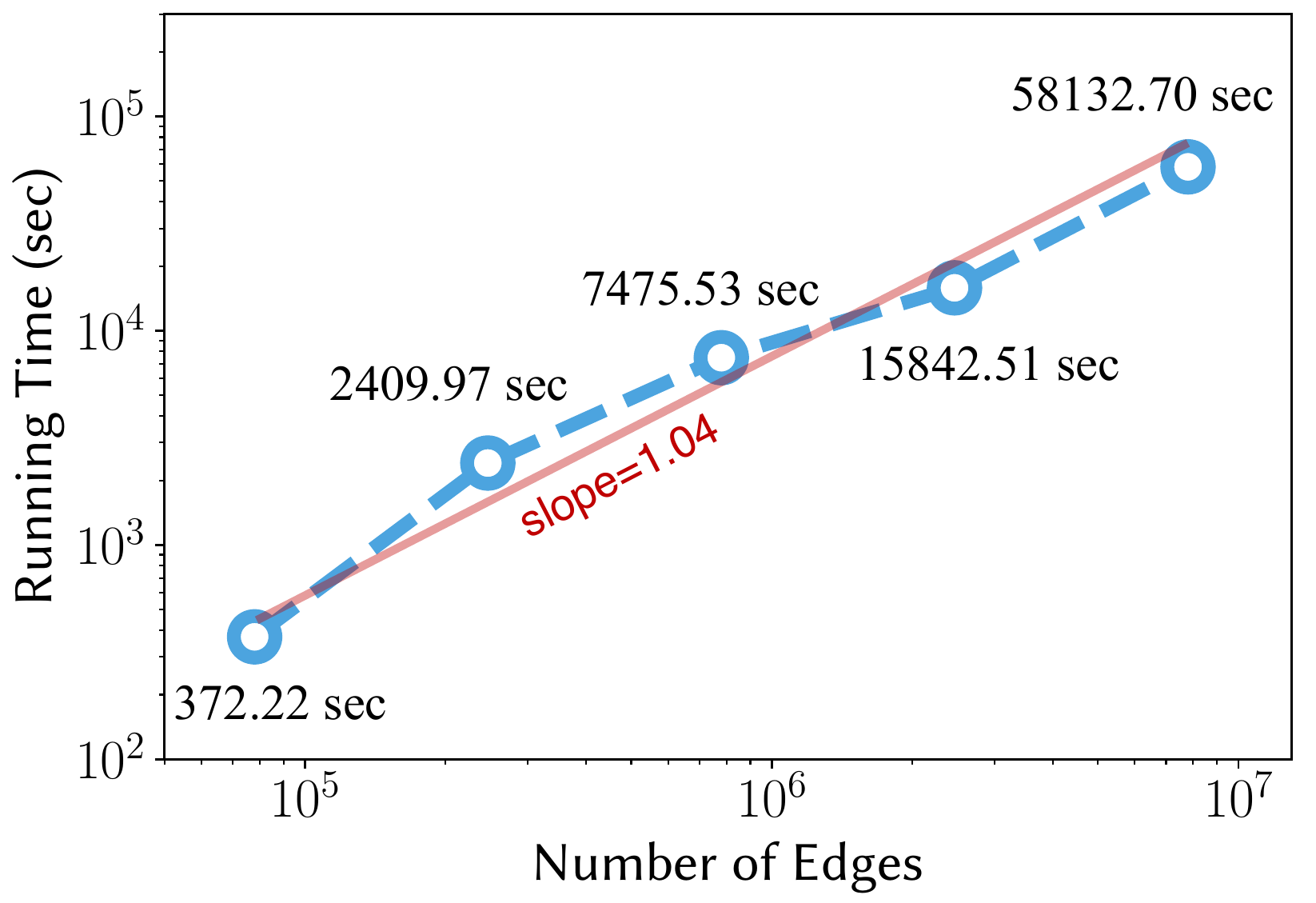}\\
	\caption{\method provides near-linear scalability with regard to the number of edges.}
	\label{fig:scalability}
\end{figure}

We measure the running times of \method varying the number of edges for scalability experiments.
{
We generate 4 synthetic subgraphs using Forest Fire Sampling (FFS)~\cite{FFS}, where
\textit{(\# of nodes, \# of edges)} are $(34203, 78000)$, $(62339, 246480)$, $(109764, 780000)$, $(162759, 2464800)$ and $(249455, 7885487)$.
Note that FFS enables generated subgraphs to follow the properties of real world graphs: heavy-tailed distributions, densification law, effective shrinking diameters~\cite{FFS}.
}
As shown in Fig.~\ref{fig:scalability},
\method scales near-linearly with the number of edges. % with the slope 1.04.

%\begin{table}[t!]
%	\caption{To evaluate the scalability, we generate $4$ graphs.
%%	in $\mathbb{R}^{t \times c}$ where $t$ corresponds to total length (time), and $c$ corresponds to the number of time series.
%	}
%	\centering
%	\label{tab:scala}
%		\resizebox{0.3\textwidth}{!}{
%	\begin{tabular}{lrrr}
%		\toprule
%		\textbf{Dataset} & \textbf{\# of nodes} & \textbf{\# of edges}  \\
%		\midrule
%		 GraphA  & $34,203$ & $78,000$ \\
%		 GraphB & $109,764$ & $780,000$   \\
%		 GraphC & $249,455$ & $7,800,000$  \\
%		\bottomrule
%	\end{tabular}}
%\end{table}

%\begin{figure*}[!t]
%	\centering
%	 	 \subfloat[Example for stars]{\includegraphics[width=0.3\textwidth]{FIG/Example.pdf}\label{fig:example1}}
%	 	 \subfloat[Example for bipartite cores]{\includegraphics[width=0.65\textwidth]{FIG/Example2.pdf}\label{fig:example_segement_bc}}\\
%	\caption{Example of a comparison between a large inconsistent structure and two small consistent structures.	}
%	\label{fig:example}
%\end{figure*}

\section{Related Work}

We present related works on four main categories: minimal description length, graph compression and summarization, hierarchical graph, and game data mining.
%Table ? presents a visual comparison of \method with alternative approaches.

\textbf{Minimal description length.}
MDL ~\cite{rissanen1983universal} principle is a model selection method, where the best model is considered the one that gives the best lossless compression.
Its use for compression ~\cite{faloutsos2007data} is related to summarization and pattern discovery.
\method exploits the MDL principle for summarizing a heterogeneous graph with hierarchical node labels.
%Even though compression is not our goal, from an MDL's standpoint; a good compression is a good summary, becoming an essential tool for summarization.

\textbf{Graph compression and summarization.}
Works like eigendecomposition ~\cite{shah2014spotting}, modularity-based optimization ~\cite{blondel2008fast}, and cross association ~\cite{chakrabarti2004fully} find dense structures, such as cliques and bipartite cores.
SUBDUE~\cite{CookH94} discovers the best substructure that compresses a graph with node labels based on the MDL principle.
Slashburn~\cite{KangF11} reorders nodes of a graph to efficiently compress the graph by exploiting the characteristics of real-world graphs.
{
\cite{vcebiric2018compact} summarizes RDF graphs and
\cite{Al-DhelaanA15} performs graph summarization for hashtag recommendation on social graphs.
}
VoG~\cite{KoutraKVF14} leverages the MDL principle to summarize a homogeneous graph with common structures.
Shah et al.~\cite{ShahKZGF15} propose TimeCrunch that extends VoG method for dynamic graphs.
{
None of the above methods target a heterogeneous MMORPG graph, a heterogeneous graph with hierarchical labels, which \method focuses on.
}

\textbf{Hierarchical graph.}
Hierarchical graphs have been studied for node embedding and classification.
To effectively learn node embeddings in a graph, GNN methods use hierarchical structures.
HGP-SL~\cite{zhang2019hierarchical} and DiffPool~\cite{NIPS2018_7729} improve the performance of predicting unknown graph labels using hierarchical representations of graphs.
StructureNet ~\cite{mo2019structurenet} proposes a generator using hierarchical graph, which produces diverse and realistic mesh geometries.
Zhang et al.~\cite{multilabelClass} use hierarchical information of labels to improve multi-label classification performance.
Wendt et al.~\cite{hlp} propose a hierarchical label propagation algorithm for document classification.
Our work is the first one to summarize a heterogeneous graph with hierarchical labels.
Unlike \method, none of the above methods address the problem of hierarchical graph summarization.

\textbf{Game data mining.}
There have been several works for analyzing game data.
%Bauckhage et al.~\cite{behaviorclustering} introduce several clustering algorithms to analyze the behaviors of players.
Drachen et al.~\cite{cluster2} analyze the behaviors of players in MMORPGs using a clustering approach.
Bernardi~\cite{botdetection} propose a bot detection algorithm by analyzing the behaviors of players.
Yang et al.~\cite{clusteringGame} cluster game players by analyzing their purchase records.
Thompson et al.~\cite{sentiment} analyze chat messages of players in StarCraft 2 using a lexicon-based approach.
We represent a popular MMORPG data as a heterogeneous graph, and analyze the graph for summarization.

\section{Conclusion}

In this paper, we analyze a massive real-world MMORPG graph using hierarchical graph summarization.
For the purpose, we propose \method, a novel graph summarization algorithm for a heterogeneous graph with hierarchical node labels.
Based on the MDL principle,
\method decomposes a heterogeneous graph into subgraphs, identifies each subgraph as a structure, segments each structure by considering hierarchical labels of its nodes,
and
summarizes the graph.
Through extensive experiments on a large real-world MMORPG graph,
we show that \method is a useful and scalable tool for graph
summarization.
Thanks to \method,
we find key structures and similar users in the MMORPG graph.

%\section{References}
\section*{Acknowledgement}
This research is supported from NCSOFT Co.

\bibliography{mybibfile}

\end{document}